%% file: template.tex
\documentclass{vgtc}                          

\input{packages}

\onlineid{1011}

\vgtccategory{Research}

\vgtcinsertpkg

\title{Distributed Path Compression for Piecewise Linear Morse-Smale Segmentations and Connected Components}

\author{Michael Will\thanks{e-mail: mswill@rptu.de}\\ %
        \scriptsize RPTU Kaiserslautern-Landau %
\and Jonas Lukasczyk\thanks{e-mail: lukasczyk@rptu.de}\\ %
     \scriptsize RPTU Kaiserslautern-Landau %
\and Julien Tierny\thanks{e-mail: julien.tierny@sorbonne-universite.fr}\\ %
     \scriptsize CNRS and Sorbonne Universit\'{e} %
\and Christoph Garth\thanks{e-mail: garth@rptu.de}\\ %
     \scriptsize RPTU Kaiserslautern-Landau %
}

\abstract{
    This paper describes the adaptation to a distributed computational setting of a well-scaling parallel algorithm for computing Morse-Smale segmentations based on path compression.
    Additionally, we extend the algorithm to efficiently compute connected components in distributed structured and unstructured grids, based either on the connectivity of the underlying mesh or a feature mask.
    Our implementation is seamlessly integrated with the distributed extension of the Topology ToolKit (TTK), ensuring robust performance and scalability.
    To demonstrate the practicality and efficiency of our algorithms, we conducted a series of scaling experiments on large-scale datasets, with sizes of up to $4096^3$~vertices on up to 64~nodes and 768~cores.
} 


\begin{document}


\input{sections/introduction}
\input{sections/related}

\input{sections/background}

\input{sections/methods}
\input{sections/experiments}
\input{sections/conclusion}

\acknowledgments{
This work is partially supported by the European Commission grant ERC-
2019-COG “TORI” (ref. 863464, https://erc-tori.github.io/).}

\bibliographystyle{abbrv-doi}

\bibliography{template}
\end{document}

%% file: packages.tex
\graphicspath{{figs/}{figures/}{pictures/}{images/}{./}} 
\usepackage[utf8]{inputenc}

\usepackage{tabu}                      
\usepackage{booktabs}                  
\usepackage{lipsum}                    
\usepackage{mwe}       
\usepackage{graphicx}
\usepackage{placeins}
\usepackage{mathptmx}                  

\usepackage[table,svgnames]{xcolor}
\usepackage{framed}
\usepackage{tcolorbox}
\usepackage{multicol}
\usepackage{multirow}
\usepackage{colortbl}
\usepackage{tabularx}
\usepackage{xr}
\usepackage{todonotes}
\newcommand{\lv}{{\color{grayC}\vrule}}

\newcommand{\lh}{\arrayrulecolor{grayC}\hline\arrayrulecolor{black}}
\usepackage[export]{adjustbox}

\usepackage{tikz}
\usetikzlibrary{arrows,arrows.meta,patterns,calc,shapes,patterns.meta}
\usepackage{etoolbox}
\usepackage{pgfplots}
\DeclareUnicodeCharacter{2212}{−}
\usepgfplotslibrary{groupplots,dateplot}
\pgfplotsset{compat=newest}
\usepackage{ifthen}
\makeatletter
\patchcmd\@hex@@Hex{f\else}{F\else}{}{}
\makeatother
\usepackage{rotating}
\input{tikzUtils}

\definecolor{paperRed}{rgb}{0.86, 0.08, 0.24}
\definecolor{paperBlue}{HTML}{309dcf}

\usepackage[ruled,vlined,linesnumbered]{algorithm2e}
\usepackage{algpseudocode}
\SetVlineSkip{1pt}

\newcommand{\algorithmHSpaceValue}{\algorithmHSpaceDefault}
\newcommand{\algorithmHSpace}{\hspace*{\algorithmHSpaceValue}}
\SetAlCapHSkip{0em}
\newlength{\maxwidth}
\DontPrintSemicolon
\newcommand{\algAM}[2]%
{\makebox[\maxwidth][l]{$#1$}\algorithmHSpace$\leftarrow\;#2$}
\newcommand{\algAT}[2]%
{\makebox[\maxwidth][l]{$#1$}\algorithmHSpace$\leftarrow\;$#2}
\SetKwRepeat{Do}{do}{while}
\SetArgSty{textnormal}
\SetAlgoSkip{}
\newcommand{\algcomment}[2]{\hspace{#1em}{\color{paperBlue}\text{// }\textit{#2}}\hspace*{-5em}}
\newcommand{\algpragma}[1]{\text{\color{paperRed}\#~#1}\;}

\newcommand{\rev}[1]{#1}

\SetKwInOut{Input}{\hspace*{-0em}Inputs\hspace*{-0.5em}}
\SetKwInOut{Output}{\hspace*{-0.8em}Outputs\hspace*{-0.5em}}
\SetKwFor{ParallelFor}{parallel foreach}{do}{end}
\SetKwBlock{Parallel}{parallel}{end}
\SetInd{0.4em}{0.6em}

\newcolumntype{H}{@{}>{\lrbox0}l<{\endlrbox}}

\makeatletter
\definecolor{codebg}{cmyk}{0,0,0,0.03}
\def\@algocf@pre@ruled{\begin{tcolorbox}[colback=codebg,arc=0.2em,boxsep=0pt,left=0pt, right=0pt, top=0.5em, bottom=0.5em,boxrule=0.1mm]}%
\def\@algocf@post@ruled{\end{tcolorbox}\vspace*{-1.5em}}%
\renewcommand{\fnum@algocf}{\hspace*{0.5em}\AlCapSty{\AlCapFnt\algorithmcfname} \arabic{algocf}}
\makeatother

\DeclareMathAlphabet{\mathcal}{OMS}{cmsy}{m}{n}

\input{notations}

\usepackage{amsfonts}
\usepackage{amsmath}
\usepackage{amssymb}

\newcommand{\argmax}{\arg\!\max}

\ifpdf
  \pdfoutput=1\relax                   
  \usepackage{graphicx}                
  \DeclareGraphicsExtensions{.pdf,.png,.jpg,.jpeg} 
\else
  \usepackage{graphicx}                
  \DeclareGraphicsExtensions{.eps}     
\fi%

\graphicspath{{figures/}{pictures/}{images/}{./}} 

\usepackage{microtype}                 
\PassOptionsToPackage{warn}{textcomp}  
\usepackage{textcomp}                  
\usepackage{times}                     
\usepackage{cite}                      

%% file: tikzUtils.tex
\newcommand{\boundellipse}[3]
{(#1) ellipse (#2 and #3)}

\definecolor{branchColor1}{HTML}{1F77B4} 
\definecolor{branchColor2}{HTML}{AEC7E8} 
\definecolor{branchColor3}{HTML}{2CA02C} 
\definecolor{branchColor4}{HTML}{98DF8A} 
\definecolor{branchColor7}{HTML}{D62728} 
\definecolor{branchColor5}{HTML}{FFBB78} 
\definecolor{branchColor6}{HTML}{FF7F0E} 
\definecolor{branchColor8}{HTML}{9C00FF} 
\definecolor{branchColor9}{HTML}{D189FF} 

\definecolor{grayE}{HTML}{EEEEEE}
\definecolor{grayD}{HTML}{DDDDDD}
\definecolor{grayC}{HTML}{CCCCCC}
\definecolor{grayB}{HTML}{BBBBBB}
\definecolor{grayA}{HTML}{AAAAAA}
\definecolor{gray9}{HTML}{999999}
\definecolor{gray8}{HTML}{888888}
\definecolor{gray7}{HTML}{777777}
\definecolor{gray6}{HTML}{666666}
\definecolor{gray5}{HTML}{555555}
\definecolor{gray4}{HTML}{444444}
\definecolor{gray3}{HTML}{333333}
\definecolor{gray2}{HTML}{222222}
\definecolor{gray1}{HTML}{111111}

\definecolor{filtered}{HTML}{BBBBBB}
\definecolor{filtered2}{HTML}{CCCCCC}
\definecolor{filtered3}{HTML}{F5F5F5}

\definecolor{lm2}{HTML}{6FB0E7}
\definecolor{lm1}{HTML}{2484D6}
\definecolor{l1}{HTML}{BD0026}
\definecolor{l2}{HTML}{F03B20}
\definecolor{l3}{HTML}{FD8D3C}
\definecolor{l4}{HTML}{FED976}
\definecolor{l5}{HTML}{FFFFB2}

\definecolor{colorBrewerC12_0}{HTML}{a6cee3}
\definecolor{colorBrewerC12_1}{HTML}{1f78b4}
\definecolor{colorBrewerC12_2}{HTML}{b2df8a}
\definecolor{colorBrewerC12_3}{HTML}{70bc6b}
\definecolor{colorBrewerC12_4}{HTML}{fb9a99}
\definecolor{colorBrewerC12_5}{HTML}{eb5e60}
\definecolor{colorBrewerC12_6}{HTML}{fdbf6f}
\definecolor{colorBrewerC12_7}{HTML}{bbbbbb}
\definecolor{colorBrewerC12_8}{HTML}{cab2d6}
\definecolor{colorBrewerC12_9}{HTML}{9677b8}
\definecolor{colorBrewerC12_10}{HTML}{ffff99}
\definecolor{colorBrewerC12_11}{HTML}{b15928}
\definecolor{colorBrewerC12_12}{HTML}{595959}
\definecolor{colorBrewerC12_13}{HTML}{e80500}

\definecolor{colorBrewerS9_0}{HTML}{fff5f0}
\definecolor{colorBrewerS9_1}{HTML}{fee0d2}
\definecolor{colorBrewerS9_2}{HTML}{fcbba1}
\definecolor{colorBrewerS9_3}{HTML}{fc9272}
\definecolor{colorBrewerS9_4}{HTML}{fb6a4a}
\definecolor{colorBrewerS9_5}{HTML}{ef3b2c}
\definecolor{colorBrewerS9_6}{HTML}{cb181d}
\definecolor{colorBrewerS9_7}{HTML}{a50f15}
\definecolor{colorBrewerS9_8}{HTML}{67000d}
\definecolor{colorBrewerS9_9}{HTML}{000000}

\definecolor{ng3dt1}{HTML}{3182bd}
\definecolor{ng3dt2}{HTML}{6baed6}
\definecolor{ng3dt3}{HTML}{9ecae1}

\definecolor{ng3dn1}{HTML}{cc0000}
\definecolor{ng3dn2}{HTML}{d96666}
\definecolor{ng3dn3}{HTML}{ffabab}

\tikzstyle{contourNumber0} = [fill=white, draw=black, circle, inner sep=0pt]
\tikzstyle{contourNumber1} = [fill=white, draw=black, circle, inner sep=0.5pt]
\tikzstyle{contourNumber2} = [fill=white, draw=black, circle, inner sep=0.04em]

\tikzstyle{veryThinEdge} = [line width=0.01cm]
\tikzstyle{thinEdge} = [line width=0.03cm]
\tikzstyle{mediumEdge} = [line width=0.04cm]
\tikzstyle{thickEdge} = [line width=0.1cm]
\tikzstyle{dashedEdge} = [dashed, gray]

\tikzstyle{thickNode} = [shape=circle,draw=black,fill=black, inner sep=0pt, minimum size=4]
\tikzstyle{filteredNode} = [shape=circle,draw=filtered,fill=filtered, inner sep=0pt, minimum size=4]

\tikzstyle{arrow} = [-{Stealth[scale=1.5]}]
\tikzstyle{arrow2} = [-{Stealth[scale=1.1]}]

\definecolor{pipelineColor0}{HTML}{b3cde3}
\definecolor{pipelineColor1}{HTML}{fbb4ae}
\definecolor{pipelineColor2}{HTML}{ccebc5}

\definecolor{rLevel0}{HTML}{a93838}
\definecolor{rLevel1}{HTML}{e3aeae}
\definecolor{bLevel0}{HTML}{3e7fa4}
\definecolor{bLevel1}{HTML}{b1cfe1}
\definecolor{gLevel0}{HTML}{3ea440}
\definecolor{gLevel1}{HTML}{b1e1b2}

\definecolor{colorTheme_00}{HTML}{a71d44}
\definecolor{colorTheme_01}{HTML}{dd6767}

\definecolor{colorTheme_10}{HTML}{777777}
\definecolor{colorTheme_11}{HTML}{bbbbbb}

\definecolor{colorTheme_20}{HTML}{f27d00}
\definecolor{colorTheme_21}{HTML}{f1b473}
\definecolor{colorTheme_22}{HTML}{edc293}
\definecolor{colorTheme_23}{HTML}{f5cca1}

\definecolor{colorTheme_30}{HTML}{3ea440}

\definecolor{colorTheme_40}{HTML}{1a769c}
\definecolor{colorTheme_41}{HTML}{83a6c1}

\definecolor{simplexBG}{HTML}{dcdcdc}

\tikzstyle{simplexVertex}=[draw=black, thinEdge, fill=white, shape=circle, minimum size=7pt, inner sep=2pt]
\tikzstyle{highlightedSimplexVertexR}=[draw=colorTheme_00, line width=0.75mm, fill=colorTheme_01, shape=circle, minimum size=1.3em, inner sep=2pt]
\tikzstyle{highlightedSimplexVertexB}=[draw=colorTheme_40, line width=0.75mm, fill=colorTheme_41, shape=circle, minimum size=1.3em, inner sep=2pt]
\tikzstyle{highlightedSimplexVertexG}=[draw=colorTheme_10, line width=0.75mm, fill=colorTheme_11, shape=circle, minimum size=1.3em, inner sep=2pt]

\tikzstyle{simplexEdge}=[draw=black, line width=0.2mm]
\tikzstyle{highlightedSimplexEdgeR}=[draw=colorTheme_00, thickEdge]
\tikzstyle{highlightedSimplexEdgeB}=[draw=colorTheme_40, thickEdge]
\tikzstyle{highlightedSimplexEdgeG}=[draw=colorTheme_10, thickEdge]

\tikzstyle{simplexTriangle}=[fill=simplexBG, draw=black, line width=0.1mm]
\tikzstyle{highlightedSimplexTriangleR}=[fill=rLevel1, draw=black, line width=0.1mm]
\tikzstyle{highlightedSimplexTriangleB}=[fill=bLevel1, draw=black, line width=0.1mm]
\tikzstyle{highlightedSimplexTriangleG}=[fill=gLevel1, draw=black, line width=0.1mm]

%% file: notations.tex
\newcommand{\domain}{\mathcal{K}}
\newcommand{\vertices}[1]{\mathcal{V}(#1)}
\newcommand{\edges}[1]{\mathcal{E}(#1)}

\newcommand{\range}{\mathbb{R}}

\newcommand{\sfield}{f}
\newcommand{\mask}{m}

\newcommand{\desManifold}{d}

\newcommand{\maxima}{\mathcal{M}}
\newcommand{\neighbors}[2]{\mathcal{N}(#1,#2)}

\newcommand{\sublevelset}[1]{{#1}^{-1}_{-\infty}}

\newcommand{\Link}{Lk}
\newcommand{\simplex}{\sigma}

\newcommand{\Index}{\mathcal{I}}

\newcommand{\edge}[2]{\langle#1,#2\rangle}

%% file: sections/introduction.tex
\firstsection{Introduction}
\maketitle
Topological Data Analysis (TDA) has become a popular tool for capturing the inherent structure and features of interest of scalar field data.
It has been used for a multitude of visualization and analysis tasks, for example in the areas of fluid and combustion dynamics\cite{landge_-situ_2014, kasten_two-dimensional_2011, nauleau_topological_2022}, climate science\cite{doraiswamy_exploration_2013,hotz_topology-based_2021,nilsson_exploring_2022} or astrophysics\cite{sousbie_persistent_2011,shivashankar_felix_2016} and more~\cite{heine_survey_2016,yan_scalar_2021}.
Topological abstractions capture the global structure of data and allow researchers to extract relevant features more quickly and easily. 
As dataset sizes continue to grow, relying on a single machine for computing abstractions is becoming increasingly impractical. 
This is particularly true for methods involving larger data capture, such as increasingly accurate scientific simulations or medical imaging. 
Generally, shared computation is almost always preferred to distributed computation as communication can quickly become a bottleneck and impede the speed of the computation, especially for global problems such as the computation of TDA abstractions.
However, memory limitations necessitate data distribution across multiple machines to efficiently process and analyze large-scale datasets.

One prominent topological abstraction is the Morse-Smale (MS) complex, which segments the domain into areas of similar gradient flow and has been used for the visualization of instabilities in hydrodynamic mixing layers~\cite{laney_understanding_2006}, highlighting the dark matter cosmic web~\cite{sousbie_persistent_2011}, feature extraction of combustion simulations~\cite{landge_-situ_2014}, feature tracking~\cite{nilsson_probabilistic_2023}, and many more applications.
However, the distributed computation of Morse-Smale complexes is an underdeveloped field~\cite{gyulassy_parallel_2012}, and to the best of our knowledge at the time of writing, there is no publicly available implementation.

This paper describes Distributed Path Compression (DPC), an adaption of a well-scaling shared-memory parallel algorithm for computing the MS segmentation~\cite{maack_parallel_2023} to a distributed setting.
To evaluate our implementation, we performed weak and strong scaling experiments for DPC.
Our results show that due to the global nature of the problem and the additional communication overhead, the original shared-memory parallel implementation always outperforms the distributed version when working on a similar number of threads (e.g., 1 node with 48 threads performs better than 4 nodes with 12 threads).
However, for large datasets it is often not possible to acquire singular nodes with the needed memory requirements, necessitating data distribution.
Due to the lack of reference implementations, we were not able to compare DPC against other approaches.
Thus, we consider the provision of a public implementation for future benchmarks to be a core contribution of this~work.

\begin{figure*}[ht!]
    \centering
    \includegraphics[width = 0.25\linewidth]{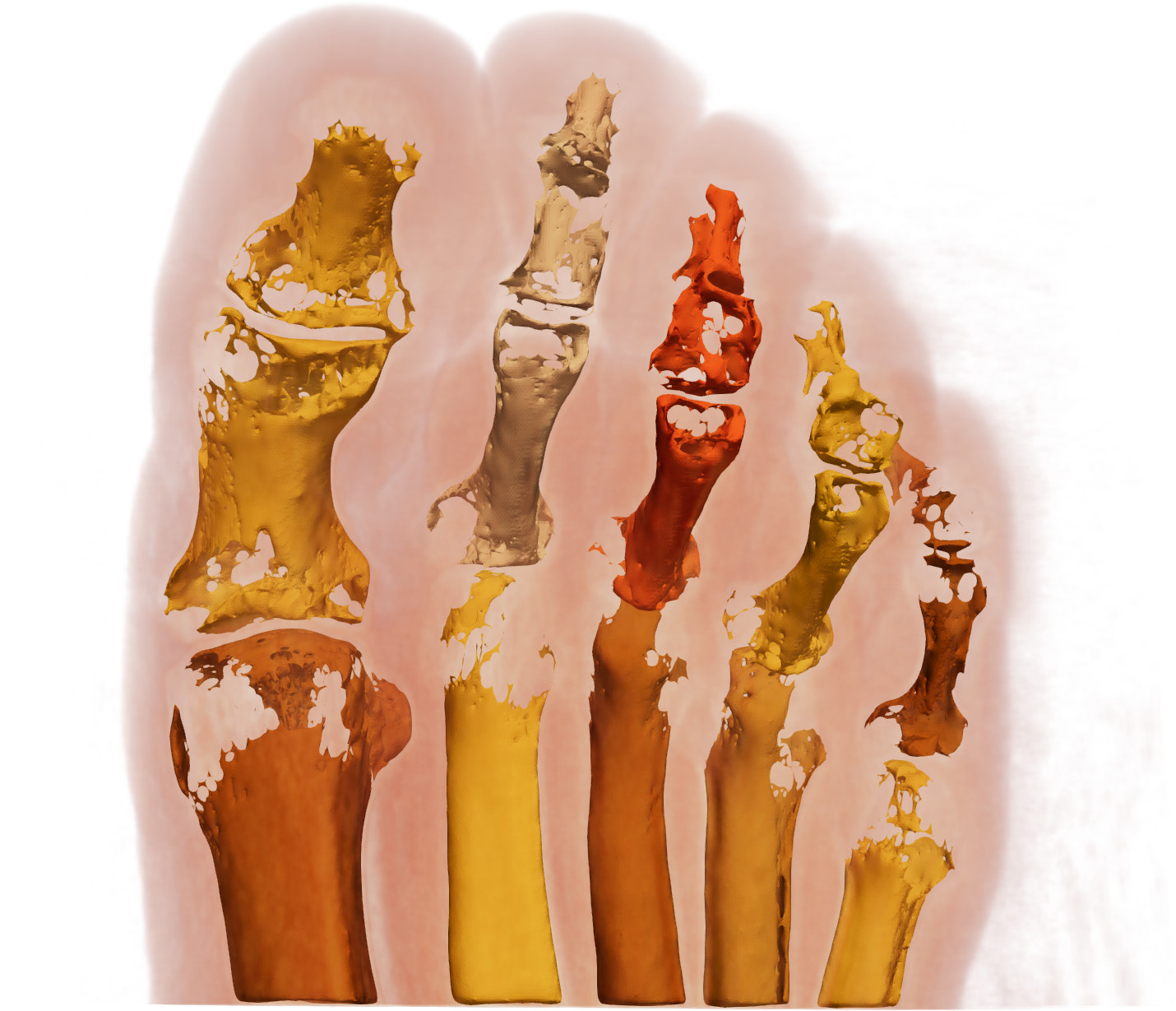}~
    \includegraphics[width = 0.4\linewidth]{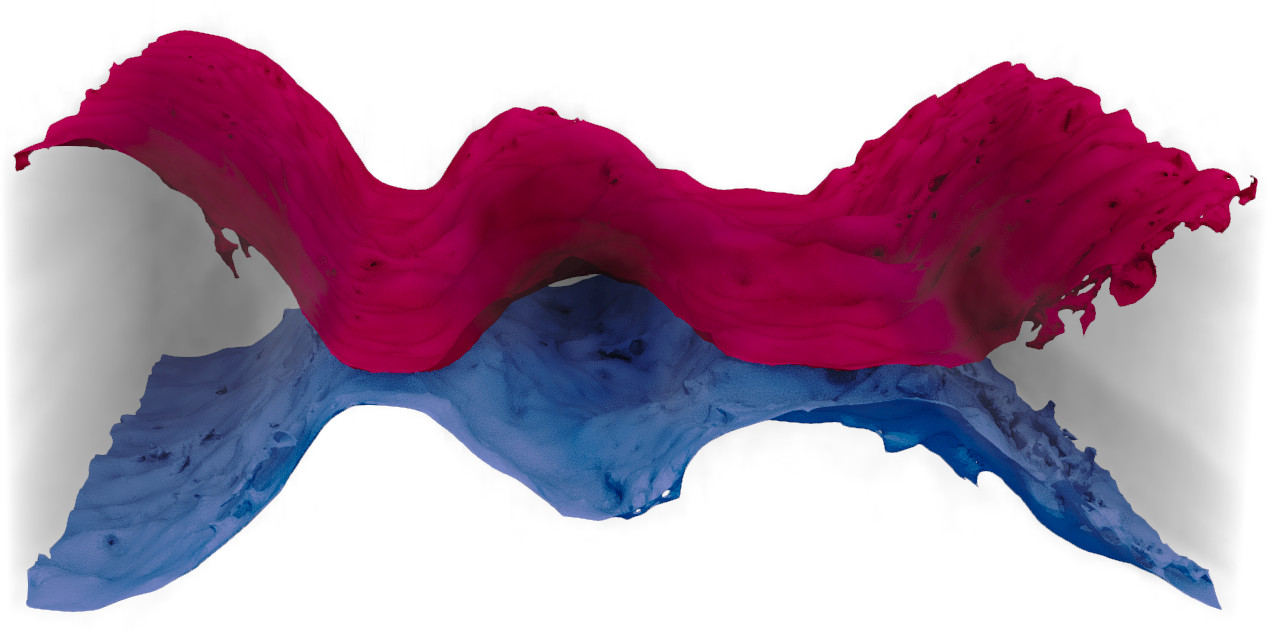}~    \includegraphics[width = 0.35\linewidth]{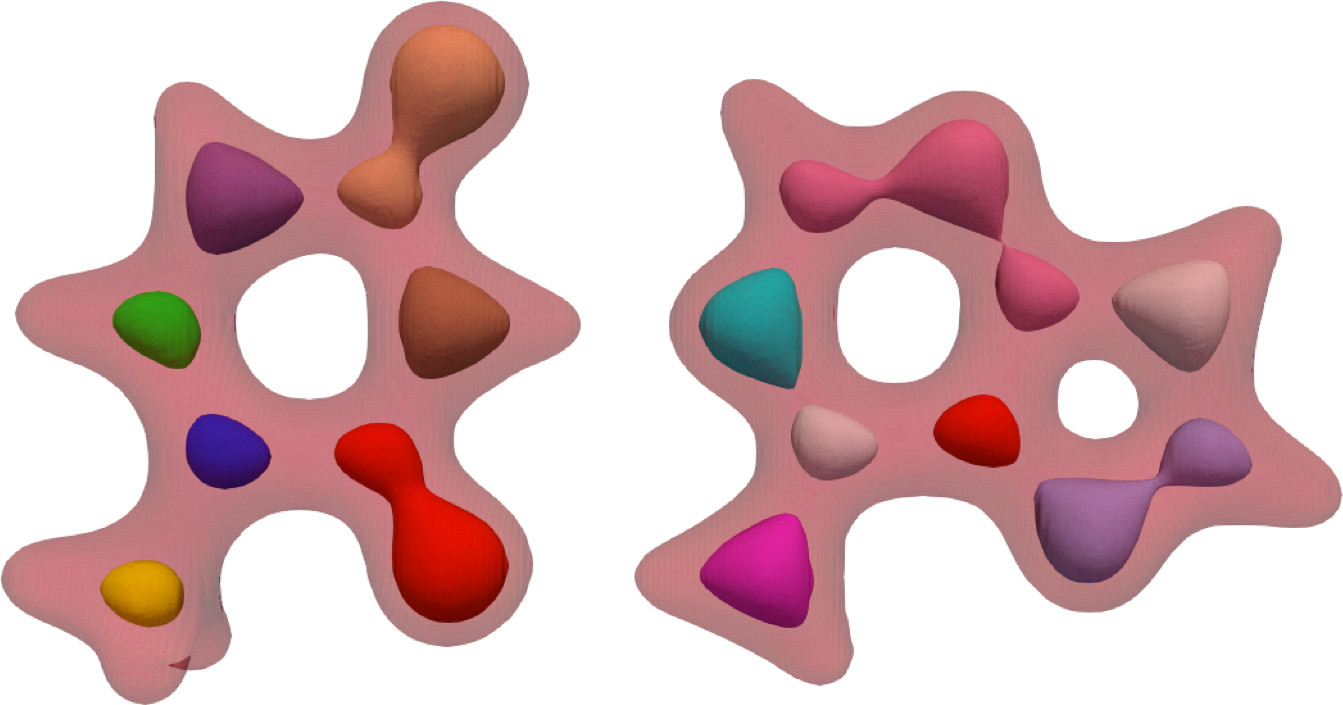}~
    \vspace*{-1em}
    \caption{
    Connected Component extraction for ctBones~\cite{ttk-data}, the magnetic reconnection~\cite{magnetic_reconnection} and the AT complex~\cite{ttk-data} datasets based on a threshold, which characterize bones of the foot, high-density boundaries and low density areas, respectively.
    Running these computations on multiple nodes allows us to use much larger datasets by using the distributed memory of all the nodes.
    }
    \label{fig:vis}
\end{figure*}

To compare DPC at least somehow to existing implementations, we also describe a modification of DPC for the computation of connected components in structured and unstructured grids~(\autoref{fig:vis}).
To this end, we compare the DPC-based connected component computation against the implementation provided in the Visualization Toolkit (VTK).
Our results show that DPC requires much less memory and performs better or, at worst, similarly to the VTK equivalent for larger node counts and problem sizes.
\vspace*{1em}

\noindent
To summarize, the main contributions of our work are:\vspace{-0.5em}
\begin{enumerate}
    \setlength\itemsep{0em}
    \item a hybrid-parallel algorithm for the computation of Morse-Smale segmentations;
    \item a hybrid-parallel algorithm for the computation of connected components based on the principle of path compression; and
    \item the integration of both algorithms into TTK for reproducibility, future benchmarks, and utilization by end-users.
\end{enumerate}

%% file: sections/related.tex
\section{Related Work}

\subsection{Morse-Smale Complex}

The Morse-Smale complex (MS complex) provides an abstract overview over the gradient flow of the scalar field~\cite{smale_generalized_1961,smale_gradient_1961}. Critical points represent areas where the gradient flow is zero and are connected by separatrices, \rev{boundary lines segmenting the domain into areas of similar flow}.
The complex was first formally defined for piecewise linear 2-manifolds by Edelsbrunner et al.\cite{edelsbrunner_hierarchical_2003} and later extended to 3-manifolds\cite{edelsbrunner_morse-smale_2003}, where the authors first derive a \emph{quasi MS complex}, which is structurally indistinguishable from the MS complex, but where the arcs may just be of \emph{a} monotone ascent or descent, not the maximal ascent or descent, to then derive the MS complex from it.
Bremer et al.\cite{bremer_topological_2004} compute the MS complex by starting from saddle points and tracing two paths of steepest ascent and two of steepest descent. These paths naturally partition the domain in the cells of the MS complex.
Gyulassy et al.\cite{gyulassy_topological_2006, gyulassy_efficient_2007} extended these ideas to 3D domains and introduced topological simplification based on the MS complex.

Gyulassy et al~\cite{gyulassy_practical_2008} describe a divide-and-conquer algorithm with on-the-fly simplification to allow for a memory efficient computation.
Their approach and the improvement done by Gyulassy in his thesis~\cite{gyulassy_combinatorial_2008} are also based on discrete Morse theory.
Robins et al.~\cite{robins_theory_2011} extended these ideas to the first provably correct algorithm for computing the MS complex using discrete Morse theory.

Gyulassy et al.~\cite{gyulassy_conforming_2014} presented a new approach for computing the MS complex, by using streamlines to compute mountains/basins, which leads to an MS complex, whose accuracy depends on the accuracy of the integration used.
Günther et al.~\cite{gunther_memory-efficient_2011,gunther_efficient_2012}, while mostly aiming to compute persistent homology, used the MS complex as an intermediate step to compute the persistence. The MS complex is extracted by integrating along critical points in parallel.
Shivashankar et al. presented algorithms for the parallel computation of 2D~\cite{shivashankar_parallel_2012} and 3D~\cite{shivashankar_parallel_2012-1} MS complexes, based on discrete Morse theory. They first compute the discrete gradient field and then extract the relevant manifolds as collections of gradient paths, by breadth-first search starting from critical points.
Gyulassy et al.~\cite{gyulassy_shared-memory_2019} presented an algorithm with improved accuracy, while still presenting substantial speedups. They first generate a version on discrete Morse theory and then modify it using the numerically traced features. 
Subhash et al.~\cite{subhash_gpu_2020} presented the first GPU based algorithm for the MS complex. One previous computational bottleneck was correctly identifying the structure of saddle-saddle connection. They compute the connectivity via a series of matrix operations which allows them to count paths completely lock free.
Gerber et al.~\cite{gerber_visual_2010} presented the Morse-Smale Approximation which is conceptually similar to our segmentation, however, they follow the steepest \emph{k}-nearest neighbors, instead of the steepest direct neighbor. 
Maack et al.~\cite{maack_parallel_2023} computed the ascending and descending segmentation of the domain using path compression on the discrete gradient field. 
These are then merged into a fast preview of the MS complex. This algorithm is the basis for the distributed version we use.

Concepts like local-global or fully distributed representations of the Morse-Smale complex are still underdeveloped. 
Gyulassy et al.~\cite{gyulassy_parallel_2012} presented a method in which the data is distributed over multiple blocks, where each block computes their local gradient and MS complex, which are subsequently merged into one complex.
However, this approach depends on partial and local simplification to make it feasible for larger complexes, and the distributed implementation is not publicly available.

\subsection{Distributed Union-Find / Connected Components}

Path Compression can be used as an efficient algorithm type for the \emph{Find} component of the Union-Find data structure. 
While Union-Find by itself is inherently sequential, there have been efforts towards parallelizing and distributing it.
In the following we will present some distributed Union-Find results and elaborate on how they differ from our approach.
While there are many other algorithms for computing connected components (such as ones based on parallel domain decomposition ~\cite{shun_simple_2014, lamm_communication-efficient_2022} or random edge sampling~\cite{gazit_optimal_1986,halperin_optimal_2001}), we solely focus on ones based on Union-Find.

Using a variation of path compression for finding connectivity in shared memory environments was first described by Shiloach and Vishkin~\cite{shiloach_ologn_1982}. Their theoretical model worked on $n+2m$ processors, with $n$ and $m$ being the number of vertices and edges respectively.
Later, Cybenko et al.\cite{cybenko_practical_1988} presented some of the first distributed parallel algorithms for computing connected components based on Union-Find, which can be seen as the basis for many of the following methods, but these algorithms exhibited poor scaling behaviour.
However, their distributed model, which is based on merging local subsets of the data until one processor has the complete solution, showed weak scaling behaviour. 
In contrast, the algorithm by Manne and Patwary~\cite{manne_scalable_2010} is computing as much local Union-Find work as possible to minimize the needed merges. 
One important distinction to our approach is that we do not care for complete Union-Find forest, but only the final segmentation labeling for each component.\\
Iverson et al.\cite{iverson_evaluation_2015} evaluated multiple algorithms for computing connected component labeling of graphs distributed across multiple processors. 
Our algorithm is a variation of the distributed Union-Find algorithm (which is in turn a variation of the original one by Shiloach and Vishkin~\cite{shiloach_ologn_1982}), which was shown by them to scale well.\\
Friederici et al.\cite{friederici_distributed_2019} used a distributed version of their previous~\cite{friederici2018efficient} shared-memory Union-Find algorithm for efficient percolation analysis in turbulent flows.
Their algorithm allows for arbitrary representatives for the components, which may lead to more efficient computation, as less pointers need to be rearranged.\\
In contrast to our method which uses one synchronous communication step and relies on a given distribution done by VTK, Xu et al.~\cite{xu_asynchronous_2021} present an asynchronous Union-Find method with dynamic redistribution for load-balancing reasons.

%% file: sections/background.tex
\section{Background}
This section provides the theoretical background of the proposed approach and introduces the notations used throughout the manuscript. For a comprehensive introduction to computational topology, we refer the reader to the textbook of Edelsbrunner and Harer~\cite{edelsbrunner_computational_2009}.

\subsection{Scalar Fields}
The input of our approach is a piecewise-linear (PL) \emph{scalar field} \mbox{$\sfield:\domain\rightarrow\range$}, where real-valued data is given at the vertices of a connected simplicial complex $\domain$, and values on edges are linearly interpolated.
We denote the vertices (0-simplices) and edges (1-simplices) of the complex $\domain$ with $\vertices{\domain}$ and $\edges{\domain}$, respectively.
Neighbor vertices of a vertex $v$ are denoted by $\neighbors{v}{\domain} = \{u\in\vertices{\domain}~:~\edge{v}{u}\in\edges{\domain}\}$.
$\domain$ does not need to be simply connected, but we require that $\sfield$ is injective on the vertices of $\domain$, which can always be enforced by applying a variant of \emph{Simulation of Simplicity}~\cite{edelsbrunner_simulation_1990}.

\subsection{Critical Points}
The sub-level set $\sublevelset{f}(w)$ of an isovalue $w \in \mathbb{R}$ is defined as $\sublevelset{f}(w) = \{p \in \mathcal{M} ~|~ f(p) < w\}$. 
As $w$ continuously increases, the topology of $\sublevelset{f}(w)$ changes at  specific vertices of $\domain$, called the \emph{critical points} of $f$.
Let $\Link^-(v)$ be the \emph{lower link} of the vertex $v$: 
$\Link^-(v) = \{\simplex \in \Link(v) ~|~ \forall u \in \simplex : f(u) < f(v)\}$.
The \emph{upper link} of $v$ is defined symmetrically:
$\Link^+(v) = \{\simplex \in \Link(v) ~|~ \forall u \in \simplex : f(u) > f(v)\}$.
A vertex $v$ is \emph{regular} if and only if both $\Link^-(v)$ and $\Link^+(v)$ are simply connected. Otherwise, $v$ is a \emph{critical vertex} of $f$~\cite{banchoff_critical_1970}.
A critical vertex $v$ can be classified by its \emph{index} $\Index(v)$, which is $0$ for minima, $1$ for $1$-saddles, $(d-1)$ for $(d-1)$-saddles and $d$ for maxima. Vertices for which the number of connected components of $\Link^-(v)$ or $\Link^+(v)$ are greater than $2$ are called \emph{degenerate saddles}.

\subsection{Ascending and Descending Manifolds}
Let $\maxima(\domain)\subset\vertices{\domain}$ be the set of maxima of $\domain$ for $\sfield$.
Then the \emph{descending manifold} is a map $\desManifold:\vertices{\domain}\rightarrow\maxima(\domain)$ that assigns to each vertex $v\in\vertices{\domain}$ the maximum $m\in\maxima(\domain)$ that would be reached by following the path starting at $v$ along the steepest ascent on $\domain$.
This path is a sequence of $n$ vertices $(v=v_1,v_2,...,v_n=m)$ where $v_{i+1} = \argmax_{u\in \neighbors{v_i}{\domain}}\sfield(u)$. 
The \emph{ascending manifold} is defined symmetrically for minima reached by following the path starting at $v$ along the steepest descent.
These paths are unambiguous since $\sfield$ is~injective.

\subsection{Distributed Model}
Our work is integrated into the Topology ToolKit (TTK)~\cite{hotz_overview_2021,tierny_topology_2018} and is making use of its distributed capabilities. 
We will briefly recap some parts in the relevant places, while referring to the introduction paper by Le Guillou et al.~\cite{guillou_generic_2023} for more detail. 
One of the most important data-structures in TTK is the triangulation, which allows for constant time traversal queries, transformation of local into global ids and reverse, extraction of the rank ids of vertices and more.
For this, a pre-processing step is necessary, in which the developer of an algorithm specifies which traversal types will be needed, so they can be pre-computed and cached.
As TTK mostly relies on analysis \emph{pipelines}, which allows for many analysis algorithms or \emph{filters} to be chained together, this allows for relevant information to be computed once and used by the whole pipeline.

%% file: sections/methods.tex
\section{Methods}
In this section, we will go over the preprocessing which needs to happen at a previous step of the pipeline; recap the algorithm from Maack et al.~\cite{maack_parallel_2023}, present our distribution process; and show how the concept of Distributed Path Compression (DPC) can be applied to compute connected components.

\subsection{Preprocessing}
We need one layer of ghost vertices for our distribution scheme to work.
While every rank knows which of its simplices are actually ghost simplices belonging to other ranks, the inverse is not true (without additional computation): a rank does not intuitively know which other ranks depend on its simplices. Therefore, we \emph{request} information from other ranks and do not preemptively \emph{supply} information.
While we do not need unique scalar values for the connected components, we need to remove ambiguity for the MS segmentation. We apply a variant of \emph{Simulation of Simplicity}\cite{edelsbrunner_simulation_1990}, implemented by TTK in the \texttt{ttkArrayPreconditioning} filter, based on globally sorting the vertices according to their scalar values while breaking ties by their global vertex id, and then creating a global order field that assigns to each vertex its corresponding index in the sorted~array.

In structured grids, the global ids and the value disambiguation can be computed on-the-fly by special data structures in TTK.
For this and the neighborhood relations, we need to precondition the triangulation with the \texttt{preconditionDistributedVertices} and \texttt{preconditionVertexNeighbors} functions of TTK.
In TTK taxonomy terms, our algorithms are in the class of data-dependent communication, because the amount of communication is dependent on the data distribution, even though we only have one communication step.
Note that there are two types of vertex ids, local ids which are unique in the ranks, and global ids over the different ranks.
For brevity, we omit detailed descriptions of when they have to be converted, but generally local ids are used for addressing arrays in the ranks and global ids are used during the communication phase.

\begin{figure*}
    \centering
    \input{figures/illustration}
    \caption{
    Illustration of the distributed path compression (DPC) procedure for one connected polyline (top left).
    The polyline has the shape of a spiral and is distributed on four ranks whose boundaries are shown by red dashed lines.
    To compute connectivity, every rank needs one layer of ghost vertices (\textbf{a)}, dashed nodes and edges).
    Note, in VTK a vertex can be a ghost vertex in multiple ranks, but every vertex belongs exclusively to a single rank, which is called the vertex owner.
    The goal of the DPC procedure is to assign to every vertex the largest vertex identifier of its connected component, here $P$.
    In the first step of DPC, every rank computes a path compression for all its non-ghost vertices (\textbf{b)}).
    For instance, after this step the $R_3$ assigns to vertex $D$ and $E$ the ghost vertex $F$, and vertex $O$ is pointing towards vertex $P$.
    For details regarding the path compression on a single rank we refer the reader to the work of Maack~et~al.~\cite{maack_parallel_2023} and the summary described in~\autoref{sec:asc_desc}.
    The next step involves a cross-rank communication in which all ghost vertices retrieve the current pointers of their owners (table, column $P_0$).
    For example, ghost vertex $A$ is owned by the $R_0$ and is currently pointing towards $B$, so the $R_1$ (which contains $A$ as a ghost vertex) retrieves this assignment.
    Next, DPC performs a path compression on the ghost vertices (table, columns $P_1, P_2, P_3$).
    Here, after three iterations all ghost vertices point towards vertex $P$, which is communicated across ranks (\textbf{c)}).
    Finally, every rank needs to perform one more iteration of a local path compression to correctly update all pointers (\textbf{d)}).
    }
    \label{fig:spiral}
\end{figure*}
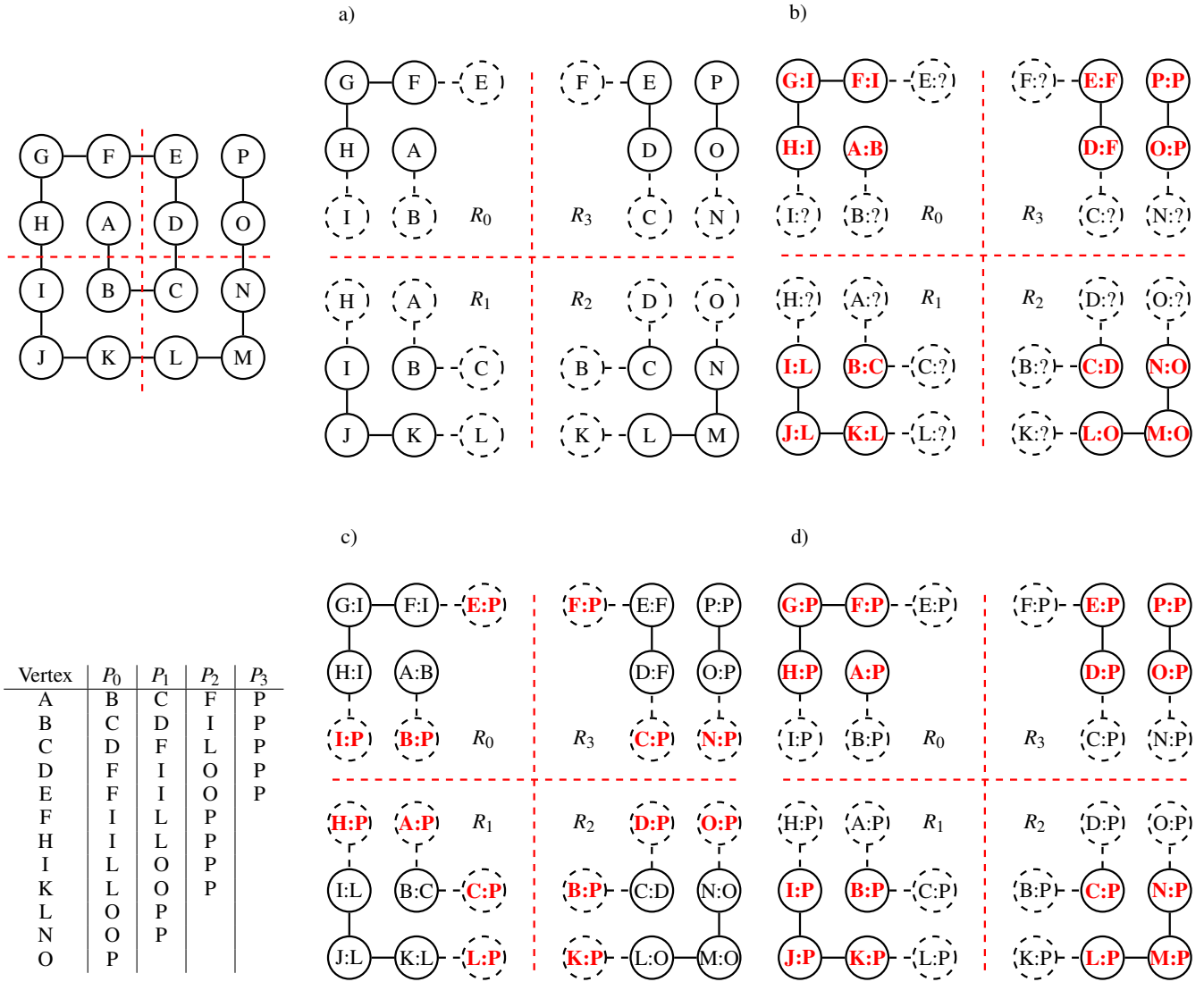

\subsection{Ascending and Descending Segmentation}
\label{sec:asc_desc}
\input{algorithms/descending_manifold_distributed}
The MS segmentations divide the scalar field into areas of similar gradient flow, therefore all vertices of which the steepest ascent and descent terminate in the same extrema, are in the same segment.
In the following we will only describe the process for the descending segmentation (steepest ascent), the process for the ascending segmentation is symmetrical.

We use \emph{path compression}, also known as pointer doubling, to efficiently compute the integral lines, as it has been shown to scale well~\cite{seidel_top-down_2005,maack_parallel_2023}.
Initially, each vertex points to its largest neighbor and then, in each global iteration, each vertex points to the vertex its current pointer pointers to. This effectively doubles the step size in each iteration. 

We outline the overall distributed algorithm in \autoref{alg:pc}.
At first, each vertex is assigned to its largest neighbor~(line~3-5).
One important change from the non-distributed to the distributed setting is seen in line 7: as previously described, we make use of ghost cells and vertices.
If a vertex is a ghost vertex, it lies on the boundary between two ranks and is present in one rank while actually belonging to another rank. 
This means that the non-owning rank has no information about the gradient behaviour in the owning rank.
Therefore, we pretend those vertices are maxima and let them point to themselves (lines 7-8).
We additionally save them separately in a vector to handle them later in the communication step.
For this, we use a C++ structure containing the vertex id, the owning rank of this vertex and the actual target to which it should point.
The target is initially set to -1 and needs to be filled by the owning rank and communicated back.

To locally perform path compression in parallel, we distribute all vertices among the available threads (lines 9-10). 
Each thread processes its own list of active vertices, where a vertex remains active as long as it is not pointing to a maximum.
For each active vertex $v$, the thread first retrieves the current pointer $u$ of $v$ (line 13), then retrieves the vertex $w$ pointed to by $u$ (lines 14-15). 
While only the current thread updates the pointer of $v$, another thread might update the pointer of $u$ during this process. 
Therefore, the first lookup does not need synchronization, but the second lookup requires an atomic read lock (line 14). 
This could be prevented by maintaining two arrays where in one iteration of path compression, you only read from one array and write into the other, and switching the arrays after each iteration.
However, this would come at higher memory usage and our experiments have shown that light synchronization outperforms the solution with two arrays.
If $u$ and $w$ are equal, indicating that $v$ now points to a maximum, $v$ can be removed from the list of active vertices (line 17). 
Otherwise, $v$'s pointer is updated to point to $w$ (line 19). 
Since only the current thread updates $v$'s pointer, this write operation does not require synchronization.

\subsection{Distributed Communication}
After computing the local segmentations per rank, we need to do one communication phase to compute the correct segmentations over all ranks, described in \autoref{alg:egv}.
For this, each rank sends ids of the needed ghost vertices with their owners to rank 0 (line 4).
Rank 0 then changes the ordering from who \emph{needs} the ranks to who \emph{owns} the ranks (line 5-8) and request this information from the actual ranks (line 10). 
Using the \emph{MPI\_Allgather} command (line 13), the actual targets for requested vertices get shared with all ranks.
We need to share this with every rank, because a vertex in the owning rank may actually point to another ghost vertex at the other end of the rank, when segmentations are stretching over multiple ranks.
Each rank now has all the information needed to build up a local ghost vertex pointer graph which can be locally compressed one last time (line 15-25).
Finally, they need to walk over their vertices and replace all the ones pointing to ghost vertices with the correct targets received and compressed earlier (line 27-33).
\input{algorithms/exchangeGhostVertices}

In our distribution approach (highlighted in \autoref{alg:egv} and used in the other algorithms), rank 0 performs additional organizational work.
Specifically, rank 0 performs a global path compression on the ghost vertices and then communicates the result to all ranks such that they can update their pointers in a single iteration.
Alternatively, this could be implemented solely based on neighbor-to-neighbor communication as opposed to one-to-all.
However, to resolve segments that stretch across multiple ranks, this approach requires multiple iterations in which ghost-cell pointers are updated between neighbors.
Additionally, this approach also introduces technical problems while resolving local and global ids, because a rank might retrieve pointers to vertices not belonging to it (neither as a ghost vertex, nor a normal vertex).
The build-in data structures in TTK do not support this case.
Our initial experiments have shown that the alternative approach incurs a higher communication overhead and requires custom data structures to resolve ids, which is why we opted to the first approach that guarantees convergence in one iteration at the expense of a global one-to-all communication.

\subsection{Connected Components}
\input{algorithms/connected_components}

Our algorithm for computing the MS segmentations can easily be adapted for computing connected components, either implicitly based on a given feature mask or explicitly based on extracted geometry. 
\rev{We describe it in \autoref{alg:cc}}.

Our algorithm for connected components follows out of our method for MS segmentations. 
Before our main algorithm, we compute a feature mask on our scalar field, which is generic and can highlight any areas of interest e.g. all vertices with a value exceeding a given threshold.
Then, we start as previously, by letting all vertices with a positive feature mask point to their largest neighbors, but with two important changes: the largest neighbor is not chosen based on the actual scalar value, but on the scalar id (therefore they can also be computed on pure geometry without any scalar data on it) and we only consider the largest neighbor for which the feature mask is also positive (line 6).
Similar to the segmentation algorithm, we let ghost vertices point to themselves and collect them for later(lines 8-9), but unlike in the segmentation algorithm, we do not need to exchange all the ghost vertices with our neighboring ranks, but only the masked ones, which can significantly improve performance.
Vertices with a non-positive feature mask immediately point to some negative value and are not considered for the further path compression steps (line 12).

\begin{figure}
    \centering
    \resizebox{0.95\columnwidth}{!}{%
        \includegraphics{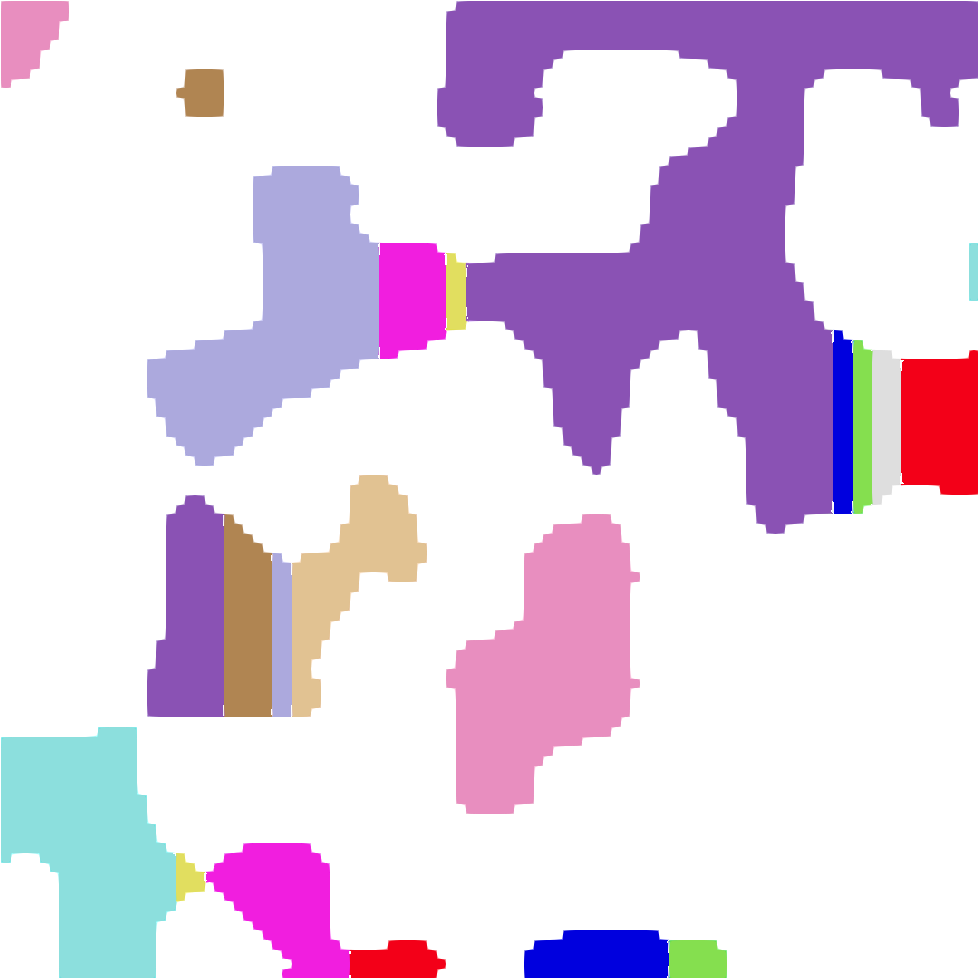}
        \hspace{3em}
        \includegraphics{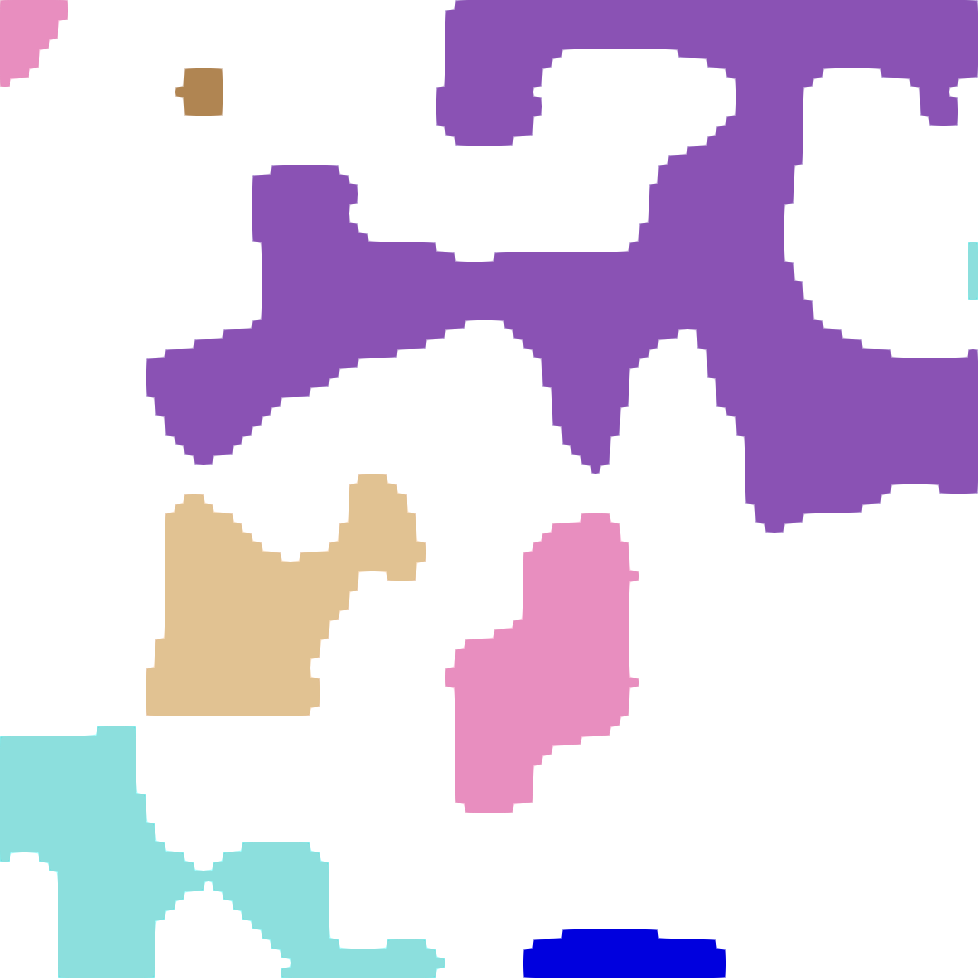}
    }\vspace*{-0.5em}
    \caption{Before (left) and after (right) applying the second pass path compression to merge the sub-segmentations in the connected components. How the segmentations are actually merged is not relevant, it just has to be done in a consistent manner. We have chosen that the segmentation whose target has a lower id gets attached to the one with the higher id. In our example dataset, the id-generation is dominated by the y-direction, therefore the remaining labeling is the one whose segment stretches furthest in positive~\mbox{x-direction}.}
    \label{fig:secondPC}
\end{figure}

We then run our path compression on the relevant vertices until convergence (lines 13-23).
These are not the correct final segmentations (as can be seen in \autoref{fig:secondPC}), as one connected component can have multiple local maxima in their id distribution.
Therefore we need to initialize a second iteration of path compression with slightly different starting conditions:
each (featured) vertex does not point to the largest neighbor, but to the neighbor with the largest pointer (line 29).
After one more path compression (lines 30-40) we have the correct local segmentation into connected components.
Finally, the distributed communication phase is the same as in the MS segmentation algorithm, the actual distribution does not care what the pointers actually signify.

%% file: figures/illustration.tex
\tikzstyle{exNode}=[draw=black, circle, line width=0.3mm, inner sep=0cm,minimum size=2em]
\tikzstyle{exEdge}=[draw=black, line width=0.3mm]
\tikzstyle{exGhost}=[dashed]
\tikzstyle{highlight}=[text=red,font=\bfseries]
\tikzstyle{exRankBoundaries}=[draw=red, dashed, line width=0.3mm]

\hfill~
\begin{tikzpicture}
    \node[exNode] (A) at (0,0) {A};
    \node[exNode] (B) at (0,-1) {B};
    \node[exNode] (C) at (1,-1) {C};
    \node[exNode] (D) at (1,0) {D};
    \node[exNode] (E) at (1,1) {E};
    \node[exNode] (F) at (0,1) {F};
    \node[exNode] (G) at (-1,1) {G};
    \node[exNode] (H) at (-1,0) {H};
    \node[exNode] (I) at (-1,-1) {I};
    \node[exNode] (J) at (-1,-2) {J};
    \node[exNode] (K) at (0,-2) {K};
    \node[exNode] (L) at (1,-2) {L};
    \node[exNode] (M) at (2,-2) {M};
    \node[exNode] (N) at (2,-1) {N};
    \node[exNode] (O) at (2,0) {O};
    \node[exNode] (P) at (2,1) {P};

    \draw[exEdge] (A)--(B)--(C)--(D)--(E)--(F)--(G)--(H)--(I)--(J)--(K)--(L)--(M)--(N)--(O)--(P);

    \draw[exRankBoundaries] (-1.5,-0.5) -- (2.5,-0.5);
    \draw[exRankBoundaries] (0.5,-2.5) -- (0.5,1.5);

    \node at (0,-3.38) {};
\end{tikzpicture}~
\hspace{0.4cm}~
\begin{tikzpicture}
    \begin{scope}[xshift=-1.25cm]
        \node at (-1,2) {a)};
        \node[exNode] (A) at (0,0) {A};
        \node[exNode,exGhost] (B) at (0,-1) {B};
        \node[exNode,exGhost] (E) at (1,1) {E};
        \node[exNode] (F) at (0,1) {F};
        \node[exNode] (G) at (-1,1) {G};
        \node[exNode] (H) at (-1,0) {H};
        \node[exNode,exGhost] (I) at (-1,-1) {I};
    
        \draw[exEdge,exGhost] (A)--(B);
        \draw[exEdge,exGhost] (E)--(F);
        \draw[exEdge,exGhost] (H)--(I);
        \node at (1,-1) {$R_0$};

        \draw[exEdge] (F)--(G)--(H);        
    \end{scope}

    \begin{scope}[xshift=1.25cm]
        \node[exNode,exGhost] (C) at (1,-1) {C};
        \node[exNode] (D) at (1,0) {D};
        \node[exNode] (E) at (1,1) {E};
        \node[exNode,exGhost] (F) at (0,1) {F};
        \node[exNode,exGhost] (N) at (2,-1) {N};
        \node[exNode] (O) at (2,0) {O};
        \node[exNode] (P) at (2,1) {P};   

        \node at (0,-1) {$R_3$};

        \draw[exEdge,exGhost] (C)--(D);
        \draw[exEdge,exGhost] (E)--(F);
        \draw[exEdge,exGhost] (N)--(O);
        \draw[exEdge] (D)--(E);
        \draw[exEdge] (O)--(P);     
    \end{scope}
    
    \begin{scope}[xshift=-1.25cm,yshift=-2.25cm]
        \node[exNode,exGhost] (A) at (0,0) {A};
        \node[exNode] (B) at (0,-1) {B};
        \node[exNode,exGhost] (C) at (1,-1) {C};
        \node[exNode,exGhost] (H) at (-1,0) {H};
        \node[exNode] (I) at (-1,-1) {I};
        \node[exNode] (J) at (-1,-2) {J};
        \node[exNode] (K) at (0,-2) {K};
        \node[exNode,exGhost] (L) at (1,-2) {L};

        \node at (1,0) {$R_1$};

        \draw[exEdge,exGhost] (H)--(I);
        \draw[exEdge,exGhost] (A)--(B)--(C);
        \draw[exEdge,exGhost] (K)--(L);
        \draw[exEdge] (I)--(J)--(K);
    \end{scope}
    
    \begin{scope}[xshift=1.25cm,yshift=-2.25cm]
        \node[exNode,exGhost] (B) at (0,-1) {B};
        \node[exNode] (C) at (1,-1) {C};
        \node[exNode,exGhost] (D) at (1,0) {D};
        \node[exNode,exGhost] (K) at (0,-2) {K};
        \node[exNode] (L) at (1,-2) {L};
        \node[exNode] (M) at (2,-2) {M};
        \node[exNode] (N) at (2,-1) {N};
        \node[exNode,exGhost] (O) at (2,0) {O};

        \node at (0,0) {$R_2$};
        \draw[exEdge,exGhost] (B)--(C)--(D);
        \draw[exEdge,exGhost] (K)--(L);
        \draw[exEdge,exGhost] (N)--(O);
        \draw[exEdge] (L)--(M)--(N);
    \end{scope}

    \draw[exRankBoundaries] (0.5,1.15) -- (0.5,-4.5);
    \draw[exRankBoundaries] (-2.5,-1.6) -- (3.5,-1.6);

\end{tikzpicture}~
\hspace{0.25cm}~
\begin{tikzpicture}
    \begin{scope}[xshift=-1.25cm]
        \node at (-1,2) {b)};
        \node at (1,-1) {$R_0$};

        \node[exNode,highlight] (A) at (0,0) {A:B};
        \node[exNode,exGhost] (B) at (0,-1) {B:?};
        \node[exNode,exGhost] (E) at (1,1) {E:?};
        \node[exNode,highlight] (F) at (0,1) {F:I};
        \node[exNode,highlight] (G) at (-1,1) {G:I};
        \node[exNode,highlight] (H) at (-1,0) {H:I};
        \node[exNode,exGhost] (I) at (-1,-1) {I:?};
    
        \draw[exEdge,exGhost] (A)--(B);
        \draw[exEdge,exGhost] (E)--(F);
        \draw[exEdge,exGhost] (H)--(I);
        
        \draw[exEdge] (F)--(G)--(H);        
    \end{scope}

    \begin{scope}[xshift=1.25cm]
    \node at (0,-1) {$R_3$};

        \node[exNode,exGhost] (C) at (1,-1) {C:?};
        \node[exNode,highlight] (D) at (1,0) {D:F};
        \node[exNode,highlight] (E) at (1,1) {E:F};
        \node[exNode,exGhost] (F) at (0,1) {F:?};
        \node[exNode,exGhost] (N) at (2,-1) {N:?};
        \node[exNode,highlight] (O) at (2,0) {O:P};
        \node[exNode,highlight] (P) at (2,1) {P:P};   
    
        \draw[exEdge,exGhost] (C)--(D);
        \draw[exEdge,exGhost] (E)--(F);
        \draw[exEdge,exGhost] (N)--(O);
        \draw[exEdge] (D)--(E);
        \draw[exEdge] (O)--(P);     
    \end{scope}
    
    \begin{scope}[xshift=-1.25cm,yshift=-2.25cm]
    \node at (1,0) {$R_1$};

        \node[exNode,exGhost] (A) at (0,0) {A:?};
        \node[exNode,highlight] (B) at (0,-1) {B:C};
        \node[exNode,exGhost] (C) at (1,-1) {C:?};
        \node[exNode,exGhost] (H) at (-1,0) {H:?};
        \node[exNode,highlight] (I) at (-1,-1) {I:L};
        \node[exNode,highlight] (J) at (-1,-2) {J:L};
        \node[exNode,highlight] (K) at (0,-2) {K:L};
        \node[exNode,exGhost] (L) at (1,-2) {L:?};
    
        \draw[exEdge,exGhost] (H)--(I);
        \draw[exEdge,exGhost] (A)--(B)--(C);
        \draw[exEdge,exGhost] (K)--(L);
        \draw[exEdge] (I)--(J)--(K);
    \end{scope}
    
    \begin{scope}[xshift=1.25cm,yshift=-2.25cm]
    \node at (0,0) {$R_2$};

        \node[exNode,exGhost] (B) at (0,-1) {B:?};
        \node[exNode,highlight] (C) at (1,-1) {C:D};
        \node[exNode,exGhost] (D) at (1,0) {D:?};
        \node[exNode,exGhost] (K) at (0,-2) {K:?};
        \node[exNode,highlight] (L) at (1,-2) {L:O};
        \node[exNode,highlight] (M) at (2,-2) {M:O};
        \node[exNode,highlight] (N) at (2,-1) {N:O};
        \node[exNode,exGhost] (O) at (2,0) {O:?};
    
        \draw[exEdge,exGhost] (B)--(C)--(D);
        \draw[exEdge,exGhost] (K)--(L);
        \draw[exEdge,exGhost] (N)--(O);
        \draw[exEdge] (L)--(M)--(N);
    \end{scope}

    \draw[exRankBoundaries] (0.5,1.15) -- (0.5,-4.5);
    \draw[exRankBoundaries] (-2.5,-1.6) -- (3.5,-1.6);

\end{tikzpicture}

\vspace{0.5cm}
\bigskip

\begin{tikzpicture}
    \node at (0,0) {
        \begin{tabular}{c|c|c|c|c}
            Vertex & $P_0$ & $P_1$ & $P_2$ & $P_3$ \\ 
            \hline
             A & B & C & F & P \\ 
             B & C & D & I & P \\ 
             C & D & F & L & P \\ 
             D & F & I & O & P \\ 
             E & F & I & O & P \\ 
             F & I & L & P &  \\ 
             H & I & L & P &  \\ 
             I & L & O & P &  \\ 
             K & L & O & P &  \\ 
             L & O & P &  &  \\ 
             N & O & P &  &  \\ 
             O & P &  &  & 
            
        \end{tabular}
    };
\end{tikzpicture}~
\hspace*{0.25cm}~
\begin{tikzpicture}

    \begin{scope}[xshift=-1.25cm]
        \node at (-1,2) {c)};
        \node at (1,-1) {$R_0$};

        \node[exNode] (A) at (0,0) {A:B};
        \node[exNode,exGhost,highlight] (B) at (0,-1) {B:P};
        \node[exNode,exGhost,highlight] (E) at (1,1) {E:P};
        \node[exNode] (F) at (0,1) {F:I};
        \node[exNode] (G) at (-1,1) {G:I};
        \node[exNode] (H) at (-1,0) {H:I};
        \node[exNode,exGhost,highlight] (I) at (-1,-1) {I:P};
    
        \draw[exEdge,exGhost] (A)--(B);
        \draw[exEdge,exGhost] (E)--(F);
        \draw[exEdge,exGhost] (H)--(I);
        
        \draw[exEdge] (F)--(G)--(H);        
    \end{scope}

    \begin{scope}[xshift=1.25cm]
    \node at (0,-1) {$R_3$};

        \node[exNode,exGhost,highlight] (C) at (1,-1) {C:P};
        \node[exNode] (D) at (1,0) {D:F};
        \node[exNode] (E) at (1,1) {E:F};
        \node[exNode,exGhost,highlight] (F) at (0,1) {F:P};
        \node[exNode,exGhost,highlight] (N) at (2,-1) {N:P};
        \node[exNode] (O) at (2,0) {O:P};
        \node[exNode] (P) at (2,1) {P:P};   
    
        \draw[exEdge,exGhost] (C)--(D);
        \draw[exEdge,exGhost] (E)--(F);
        \draw[exEdge,exGhost] (N)--(O);
        \draw[exEdge] (D)--(E);
        \draw[exEdge] (O)--(P);     
    \end{scope}
    
    \begin{scope}[xshift=-1.25cm,yshift=-2.25cm]
    \node at (1,0) {$R_1$};

        \node[exNode,exGhost,highlight] (A) at (0,0) {A:P};
        \node[exNode] (B) at (0,-1) {B:C};
        \node[exNode,exGhost,highlight] (C) at (1,-1) {C:P};
        \node[exNode,exGhost,highlight] (H) at (-1,0) {H:P};
        \node[exNode] (I) at (-1,-1) {I:L};
        \node[exNode] (J) at (-1,-2) {J:L};
        \node[exNode] (K) at (0,-2) {K:L};
        \node[exNode,exGhost,highlight] (L) at (1,-2) {L:P};
    
        \draw[exEdge,exGhost] (H)--(I);
        \draw[exEdge,exGhost] (A)--(B)--(C);
        \draw[exEdge,exGhost] (K)--(L);
        \draw[exEdge] (I)--(J)--(K);
    \end{scope}
    
    \begin{scope}[xshift=1.25cm,yshift=-2.25cm]
    \node at (0,0) {$R_2$};

        \node[exNode,exGhost,highlight] (B) at (0,-1) {B:P};
        \node[exNode] (C) at (1,-1) {C:D};
        \node[exNode,exGhost,highlight] (D) at (1,0) {D:P};
        \node[exNode,exGhost,highlight] (K) at (0,-2) {K:P};
        \node[exNode] (L) at (1,-2) {L:O};
        \node[exNode] (M) at (2,-2) {M:O};
        \node[exNode] (N) at (2,-1) {N:O};
        \node[exNode,exGhost,highlight] (O) at (2,0) {O:P};
    
        \draw[exEdge,exGhost] (B)--(C)--(D);
        \draw[exEdge,exGhost] (K)--(L);
        \draw[exEdge,exGhost] (N)--(O);
        \draw[exEdge] (L)--(M)--(N);
    \end{scope}

    \draw[exRankBoundaries] (0.5,1.15) -- (0.5,-4.5);
    \draw[exRankBoundaries] (-2.5,-1.6) -- (3.5,-1.6);

\end{tikzpicture}~
\hspace*{0.25cm}~
\begin{tikzpicture}

    \begin{scope}[xshift=-1.25cm]
        \node at (-1,2) {d)};
        \node at (1,-1) {$R_0$};

        \node[exNode,highlight] (A) at (0,0) {A:P};
        \node[exNode,exGhost] (B) at (0,-1) {B:P};
        \node[exNode,exGhost] (E) at (1,1) {E:P};
        \node[exNode,highlight] (F) at (0,1) {F:P};
        \node[exNode,highlight] (G) at (-1,1) {G:P};
        \node[exNode,highlight] (H) at (-1,0) {H:P};
        \node[exNode,exGhost] (I) at (-1,-1) {I:P};
    
        \draw[exEdge,exGhost] (A)--(B);
        \draw[exEdge,exGhost] (E)--(F);
        \draw[exEdge,exGhost] (H)--(I);
        
        \draw[exEdge] (F)--(G)--(H);        
    \end{scope}

    \begin{scope}[xshift=1.25cm]
    \node at (0,-1) {$R_3$};

        \node[exNode,exGhost] (C) at (1,-1) {C:P};
        \node[exNode,highlight] (D) at (1,0) {D:P};
        \node[exNode,highlight] (E) at (1,1) {E:P};
        \node[exNode,exGhost] (F) at (0,1) {F:P};
        \node[exNode,exGhost] (N) at (2,-1) {N:P};
        \node[exNode,highlight] (O) at (2,0) {O:P};
        \node[exNode,highlight] (P) at (2,1) {P:P};   
    
        \draw[exEdge,exGhost] (C)--(D);
        \draw[exEdge,exGhost] (E)--(F);
        \draw[exEdge,exGhost] (N)--(O);
        \draw[exEdge] (D)--(E);
        \draw[exEdge] (O)--(P);     
    \end{scope}
    
    \begin{scope}[xshift=-1.25cm,yshift=-2.25cm]
    \node at (1,0) {$R_1$};

        \node[exNode,exGhost] (A) at (0,0) {A:P};
        \node[exNode,highlight] (B) at (0,-1) {B:P};
        \node[exNode,exGhost] (C) at (1,-1) {C:P};
        \node[exNode,exGhost] (H) at (-1,0) {H:P};
        \node[exNode,highlight] (I) at (-1,-1) {I:P};
        \node[exNode,highlight] (J) at (-1,-2) {J:P};
        \node[exNode,highlight] (K) at (0,-2) {K:P};
        \node[exNode,exGhost] (L) at (1,-2) {L:P};
    
        \draw[exEdge,exGhost] (H)--(I);
        \draw[exEdge,exGhost] (A)--(B)--(C);
        \draw[exEdge,exGhost] (K)--(L);
        \draw[exEdge] (I)--(J)--(K);
    \end{scope}
    
    \begin{scope}[xshift=1.25cm,yshift=-2.25cm]
    \node at (0,0) {$R_2$};

        \node[exNode,exGhost] (B) at (0,-1) {B:P};
        \node[exNode,highlight] (C) at (1,-1) {C:P};
        \node[exNode,exGhost] (D) at (1,0) {D:P};
        \node[exNode,exGhost] (K) at (0,-2) {K:P};
        \node[exNode,highlight] (L) at (1,-2) {L:P};
        \node[exNode,highlight] (M) at (2,-2) {M:P};
        \node[exNode,highlight] (N) at (2,-1) {N:P};
        \node[exNode,exGhost] (O) at (2,0) {O:P};
    
        \draw[exEdge,exGhost] (B)--(C)--(D);
        \draw[exEdge,exGhost] (K)--(L);
        \draw[exEdge,exGhost] (N)--(O);
        \draw[exEdge] (L)--(M)--(N);
    \end{scope}

    \draw[exRankBoundaries] (0.5,1.15) -- (0.5,-4.5);
    \draw[exRankBoundaries] (-2.5,-1.6) -- (3.5,-1.6);

\end{tikzpicture}

%% file: algorithms/descending_manifold_distributed.tex
\begin{algorithm}
    \caption{DistributedPathCompression}
    \label{alg:pc}

    \Input{
        \hspace*{0.25em}$\bullet$ simplical complex $\domain$\newline
        \hspace*{0.4em}$\bullet$ scalar field $\sfield:\domain\rightarrow\range$
    }
    \Output{
        \hspace*{0.25em}$\circ$ descending manifold $\desManifold : \vertices{\domain} \rightarrow \maxima$ where\newline\hspace*{1em}$\maxima$ are the maxima of $\sfield$ on $\domain$
    }
    \BlankLine
    \hrule
    \BlankLine

    \renewcommand{\algorithmHSpaceValue}{1.3em}

    \algAT{d}{$array(\;|\vertices{\domain}|\;)$\algcomment{0.5}{create int array with $|\vertices{\domain}|$ entries}}\;   

    \algAT{gv}{$array()$\algcomment{0.5}{create (id, rank(id), target)-struct array}}\;
    \BlankLine    
    
    \renewcommand{\algorithmHSpaceValue}{2em}    
    
    \ParallelFor{vertex $v\in \vertices{\domain}$}{
        \uIf{$v$ belongs to the current rank}{
            \algAT{d[v]}{%
                $\argmax_{u\in \neighbors{v}{\domain}}f(u)$%
                \algcomment{0.2}{assign $v$ to largest neighbor    }%
            }
        }
        \Else{
            \algAT{d[v]}{v\algcomment{0.5}{treat ghost cell vertices as maxima}}\;
            gv.add($v$, rank($v$))\;
        }
    }
    \renewcommand{\algorithmHSpaceValue}{1em}
    \BlankLine


    \ParallelFor{thread t}{
        \algAT{A}{
            AssignVerticesToThread$(t,\;\vertices{\domain})$
        }\;    
        \While{$|A|>0$}{        
            \ForEach{vertex $v\in A$}{
                \algAM{u}{d[v]\algcomment{0.5}{current pointer of v}}\;
                \BlankLine
                
                \algpragma{atomic read} 
                \algAM{w}{d[u]\algcomment{0.5}{current pointer of u}}\;
                
                \BlankLine
                \uIf{$u=w$}{
                  \renewcommand{\algorithmHSpaceValue}{0.8em}
                  \algAT{A}{$A\setminus\{v\}$\algcomment{0.5}{delete $v$ from active vertices}}\;
                }
                \Else{
                  \renewcommand{\algorithmHSpaceValue}{1.7em}
                  \algAT{d[v]}{$w$\algcomment{0.5}{assign $w$ to $v$}}\;              
                }
            }
        }
    }
    \algcomment{0.5}{every rank is now finished with their local computation, we now need to share segmentations over the ghost vertices}\;
    ExchangeGhostVertices(K, d, gv)\;
    \Return{$d$}\;

\end{algorithm}

%% file: algorithms/exchangeGhostVertices.tex
\begin{algorithm}[h!]
    \caption{ExchangeGhostVertices}
    \label{alg:egv} 
    \Input{
        \hspace*{0.25em}$\bullet$ simplical complex $\domain$\newline
        \hspace*{0.4em}$\bullet$ some segmentation $d:\domain\rightarrow\range$, with\newline
        \hspace*{0.4em}$\bullet$ array of (id, rank(id), target)-structs gv
    }
    \Output{
        \hspace*{0.4em}$\circ$ segmentation $d:\domain\rightarrow\range$, with rank boundaries correctly resolved
    }
    \BlankLine
    \hrule
    \BlankLine
    
    \renewcommand{\algorithmHSpaceValue}{4.8em}
    \algAT{globalSize}{0}\;
    \renewcommand{\algorithmHSpaceValue}{9em}
    \algAT{allValuesFromRanks}{$array()$}\;
    Allreduce(gv.size(), globalSize, $+$)\algcomment{0.5}{each rank knows how many ghost vertices are needed}\;
    Gather(gv, 0, allValuesFromRanks)\algcomment{0.5}{Rank 0 gets all the needed ids, along with the ranks which need them}\;
    \uIf{$rankId==0$}{
        \renewcommand{\algorithmHSpaceValue}{7em}
        \algAT{neededPerRanks}{$array(array(|ranks|))$}\;
        \ForEach{struct $s \in allValuesFromRanks$ }{
            neededPerRanks[s.rank].add(s)
        }
    }
    \renewcommand{\algorithmHSpaceValue}{5em}
    \algAT{receivedIds}{$array()$}\;
    Scatter(neededPerRanks, receivedIds, 0) \algcomment{0.5}{each rank gets the information which of its ids are needed by some other rank   }\; 
     \ForEach{struct $s \in receivedIds$}{
        \renewcommand{\algorithmHSpaceValue}{3.3em}
        \algAT{s.target}{d[s.id]}\; 
    }
    Allgather(receivedIds, allValuesFromRanks)\algcomment{0.5}{every rank now know where every ghost points to in the neighboring rank   }\;
    \algcomment{0.5}{now we need one last path compression to resolve segmentations over multiple ranks}\; 
    \ParallelFor{thread t}{
        \renewcommand{\algorithmHSpaceValue}{0.75em}
        \algAT{A}{
            AssignIdsToThread$(t,\;i\in receivedIds)$
        }\;    
        \While{$|A|>0$}{        
            \ForEach{id $i\in A$}{
                \algAM{u}{receivedIds[v].target\algcomment{0.2}{current pointer of target}}\;
                \BlankLine
                
                \algpragma{atomic read} 
                \algAM{w}{receivedIds[u].target\algcomment{0.2}{current pointer of u}}\;
                
                \BlankLine
                \uIf{$u=w$}{
                  \renewcommand{\algorithmHSpaceValue}{0.8em}
                  \algAT{A}{$A\setminus\{v\}$\algcomment{0.5}{delete $v$ from active vertices}}\;
                }
                \Else{
                  \renewcommand{\algorithmHSpaceValue}{9em}
                  \algAT{receivedIds[v].target}{$w$\algcomment{0.5}{assign $w$ to $v$}}\;              
                }
            }
        }
    }
    \algcomment{0.5}{finally, replace each vertex pointing to a ghost vertex with the correct value (possible from multiple ranks away)}\;
    \ParallelFor{thread t}{
            \renewcommand{\algorithmHSpaceValue}{0.8em}
            \algAT{A}{
            AssignVerticesToThread$(t,\;\vertices{\domain})$
        }\;    
        \While{$|A|>0$}{        
            \ForEach{vertex $v\in A$}{
                \algAM{u}{d[v]\algcomment{0.5}{current pointer of v}}\;
                \uIf{u does not belong to the current rank}{
                    \renewcommand{\algorithmHSpaceValue}{1.7em}
                    \algAT{d[v]}{receivedIds[u].target}
                }
            }
        }
    }
    \Return{$d$}\;
\end{algorithm}

%% file: algorithms/connected_components.tex
\begin{algorithm}
    \caption{ComputeConnectedComponents}
    \label{alg:cc}

    \Input{
        \hspace*{0.25em}$\bullet$ simplicial complex $\domain$\newline
        \hspace*{0.4em}$\bullet$ feature mask $\mask:\domain\rightarrow\{0,1\}$
    }
    \Output{
        \hspace*{0.25em}$\circ$ segmentation $d$ of $\domain$ into connected components, with the segmentation label being the highest vertex id in the segmentation
    }
    \BlankLine
    \hrule
    \BlankLine

    \renewcommand{\algorithmHSpaceValue}{1.3em}

    \algAT{d}{$array(\;|\vertices{\domain}|\;)$\algcomment{0.5}{create int array with $|\vertices{\domain}|$ entries}}\;    
    \algAT{fv}{$array()$\algcomment{0.5}{create (id, rank(id), target)-struct array}}\;
    \BlankLine    
    
    \renewcommand{\algorithmHSpaceValue}{2em}    
    
    \ParallelFor{vertex $v\in \vertices{\domain}$}{

        \uIf{$m(v)=1$} {
            \uIf{$v$ belongs to the current rank}{
        \algAT{d[v]}{%
            $\argmax_{u\in \neighbors{v}{\domain} \And m(u)==1}id(u)$%
            \algcomment{0.2}{assign $v$ to the neighbor with the largest id, which is also part of the feature}\;
        }
        \Else{
            \algAT{d[v]}{v}\algcomment{0.5}{treat ghost cell vertices as maxima}\;
            fv.add($v$, rank($v$))\;
        }
        }\;}
        \Else{
        \algAT{d[v]}{$-1$}
        }
        
    }
    \renewcommand{\algorithmHSpaceValue}{1em}
    \BlankLine


    \ParallelFor{thread t}{
        \algAT{A}{
            AssignVerticesToThread$(t,\;\vertices{\domain})$
        }\;    
        \While{$|A|>0$}{        
            \ForEach{vertex $v\in A$}{
                \algAM{u}{d[v]\algcomment{0.5}{current pointer of v}}\;
                \BlankLine
                
                \algpragma{atomic read} 
                \algAM{w}{d[u]\algcomment{0.5}{current pointer of u}}\;
                
                \BlankLine
                \uIf{$u=w$}{
                  \renewcommand{\algorithmHSpaceValue}{0.8em}
                  \algAT{A}{$A\setminus\{v\}$\algcomment{0.5}{delete $v$ from active vertices}}\;
                }
                \Else{
                  \renewcommand{\algorithmHSpaceValue}{1.7em}
                  \algAT{d[v]}{$w$\algcomment{0.5}{assign $w$ to $v$}}\;              
                }
            }
        }
    }
    \algcomment{0}{finished first pathbcompression, now we need to stitch segments together}\;
    \renewcommand{\algorithmHSpaceValue}{2em}    
    
    \ParallelFor{vertex $v\in \vertices{\domain}$}{
        \uIf{$m(v)=1$} {
            \ForEach{$u\in \neighbors{v}{\domain}$}{
                \uIf{$d[u] > d[v]$}{
                    \renewcommand{\algorithmHSpaceValue}{3em}
                    \algAT{$d[d[v]]$}{$d[u]$}\algcomment{0.5}{the target of this segmentation will point to the target of the neighboring segmentation, such that one further path compression merges them}\;
                }
            } 
        }
    }
    \renewcommand{\algorithmHSpaceValue}{1em}
    \BlankLine


    \ParallelFor{thread t}{
        \algAT{A}{
            AssignVerticesToThread$(t,\;\vertices{\domain})$
        }\;    
        \While{$|A|>0$}{        
            \ForEach{vertex $v\in A$}{
                \algAM{u}{d[v]\algcomment{0.5}{current pointer of v}}\;
                \BlankLine
                
                \algpragma{atomic read} 
                \algAM{w}{d[u]\algcomment{0.5}{current pointer of u}}\;
                
                \BlankLine
                \uIf{$u=w$}{
                  \renewcommand{\algorithmHSpaceValue}{0.8em}
                  \algAT{A}{$A\setminus\{v\}$\algcomment{0.5}{delete $v$ from active vertices}}\;
                }
                \Else{
                  \renewcommand{\algorithmHSpaceValue}{1.7em}
                  \algAT{d[v]}{$w$\algcomment{0.5}{assign $w$ to $v$}}\;              
                }
            }
        }
    }
    
    ExchangeForeignVertices(K, d, fV)\;
    \Return{$d$}\;

\end{algorithm}

%% file: sections/experiments.tex
\section{Experiments}
Both of our algorithms are implemented within the \emph{Topology ToolKit (TTK)}~\cite{tierny_topology_2018, guillou_generic_2023} and heavily utilize its data structures. 
The experimental workloads were delivered via python pipelines executed with ParaView v5.12.1.
All experiments were run on the Elwetritsch cluster of RPTU Kaiserslautern-Landau on up to 64 nodes, with up to 12 cores per node, for a maximum of 768 cores in a hybrid distributed-shared setting.

We expect the computation of the ascending and descending segmentation to not scale well, due to the global nature of the problem. 
Our goal is therefore to \emph{be able} to compute this in a distributed setting at all.

We evaluate both of our algorithms on synthetically generated datasets of various resolutions based on one layer of \emph{Perlin Noise}~\cite{perlin_image_1985} with an amplitude of one and frequency in every dimension of 0.1. 
Additionally, our weak scaling study is run on a simulation of the electronic density in the Adenine Thymine complex (AT). 
This dataset is resampled using the pipelines from Guillou et al.~\cite{guillou_generic_2023} according to the number of nodes used.
A complete overview \rev{of} the timings can be seen in \autoref{tab:overview}.
For all algorithms we exclude any preprocessing steps (exchanging ghost cells, computing order arrays, computing feature masks, extracting geometry etc), as they would probably have to be done at some point in the pipeline for some of the MPI filters either way.

We compare our connectivity computation via Distributed Path Compression (DPC) with the existing VTK Connectivity filter, which works in a distributed setting.
It follows the same principles as computing connected components via DPC, with first running a local connectivity algorithm (a connected wave propagation), creating a graph of region connections across ranks (every rank gets this graph) and then running a connected component algorithm on this graph and relabeling to the correct ids.
However, it automatically transforms the data from structured grids to unstructured grids which may lead to extremely high memory usage, depending on the size of the region of interest.
In contrast, DPC can work on \emph{implicitly thresholded grids}, which maintains the data structure, but assigns a negative value and segmentation id to non-relevant areas.
Therefore, we always need one extra array of memory \rev{that is the same} size as the original grid and \rev{uses the same type of ids} (either 32- or 64-bit ids).
Most of the computation is done in-place, apart from the communication step.
However, the additional memory needed is bounded upwards by the amount of ghost vertices, which is negligible compared to the size of the whole dataset, for sensible dataset size / node count configurations.
If the regions of interest are sufficiently small or need to be extracted explicitly, this can easily be done post-hoc after computing them implicitly with a simple threshold operation for values larger than zero.

\input{tables/overview}

We ran into several problems with VTK and TTK when trying to run experiments with additional large-scale datasets:

\paragraph{VTK Ghost Cell Computation} When running our analysis pipelines with 16 or more nodes letting VTK compute ghost simplices became increasingly unreliable. 
When using verbose output of pvbatch one could see that \texttt{vtkDIYGhostUtilities} froze in the ``Exchanging ghost data between blocks'' step. 
Explicitly calling the Ghost Cell Generator helped in some cases, but for many datasets it still froze at that point (but not always). 
Using the AT example pipeline with resampling worked, but also only by adding an explicit Ghost Cell Generator.
We also tried saving the datasets into a more parallel friend format on fewer nodes and then processing it on more nodes, to no avail.
ParaView, the visualization software we used to call VTK, changed the GhostCellGenerator and removed the legacy generator in version 5.11.0, it is currently being investigated whether this issue also occurred in the legacy generator.

\paragraph{VTK / TTK Unstructured Grid Distribution} On more than 8 nodes, when switching from an Implicit to an Explicit Triangulation, such as when explicitly extracting all values above a specific threshold, and afterwards triangulating it, TTK would sometimes get such in an endless loop while trying to precondition this triangulation. 
This issue is known, however it is still unclear what exactly causes this issue and whether it is due to ParaView distribution or due to the TTK triangulation.

\subsection{Strong Scaling} 
\begin{figure}[h]
    \centering
    \includegraphics[width=0.49\columnwidth]{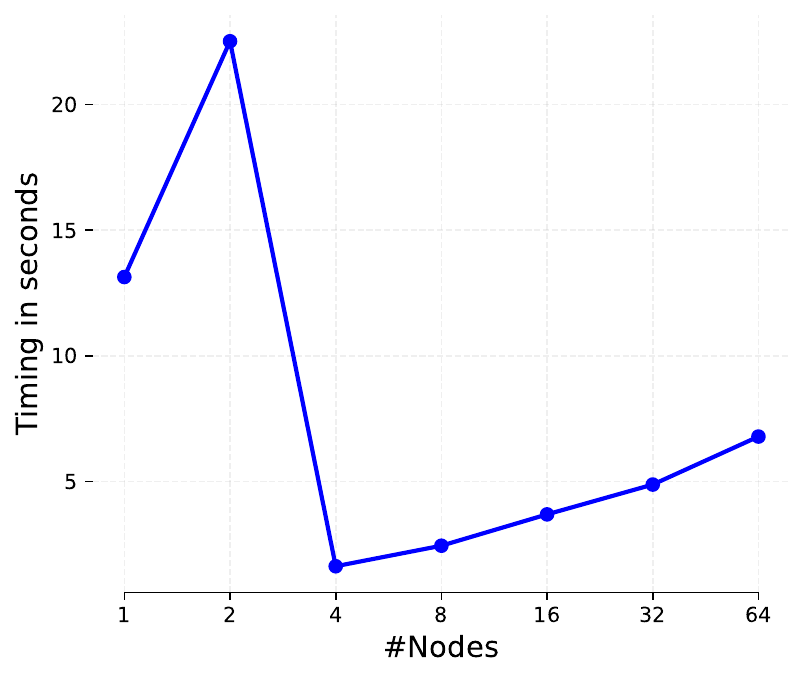}
    \includegraphics[width=0.49\columnwidth]{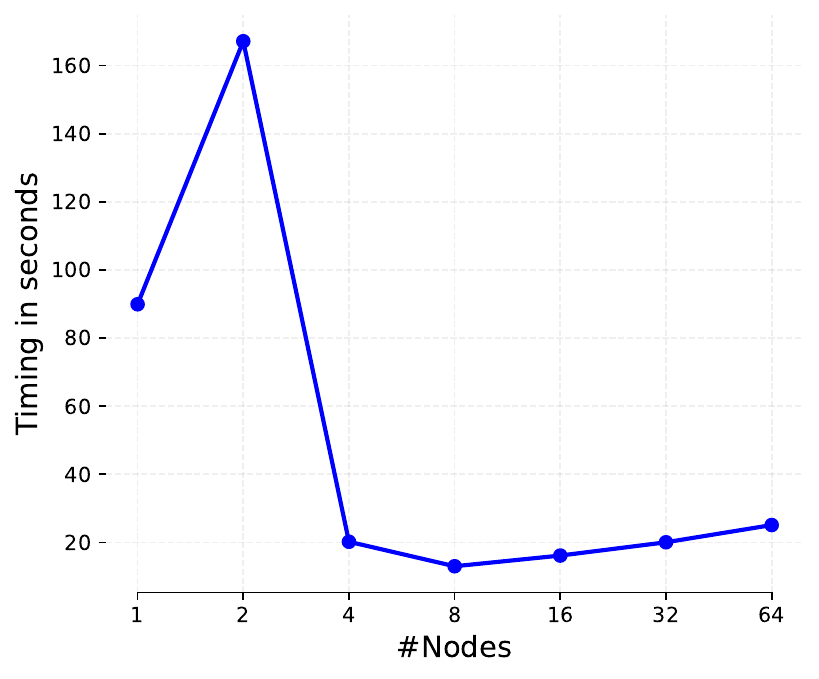}
    \caption{Timing for the strong scaling experiments of DPC based on Perlin Noise at a grid size of $512^3$ (left) and $1024^3$ (right) vertices.}
    \label{fig:pc_plot}
\end{figure}

\begin{figure}[h!]
    \centering
    \resizebox{0.95\columnwidth}{!}{%
    \input{figures/speedup_pc}
    }

    \caption{Illustration of parallel speedup and efficiency for DPC on Perlin Noise at $512^3$ \rev{(dashed blue line)} and $1024^3$ \rev{(solid blue line)}.
    The left plot shows the parallel speedup defined as the runtime of one node divided by the runtime of $n$ nodes, perfect scalability is marked with the dashed gray line.
    The right plot shows the parallel efficiency as the speedup divided by the number of nodes.
    The plot shows that the distribution step of path compression does not scale well, as more nodes significantly increase the size of the needed communication. 
    }
    \label{fig:pc_speedup}
\end{figure}
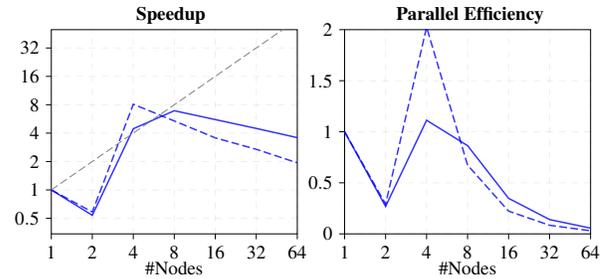

For our strong scaling study we have run the algorithms on Perlin Noise of multiple grid sizes, at $512^3$, $1024^3$, $2048^3$ and $4096^3$. 
While the larger grid was more beneficial for larger node counts, due to memory constraints only the two highest node count configurations could be run.
Therefore, we mainly focus on the large grid and compare timings starting from 4 nodes and doubling the node count up to 64 nodes.
The timings can be seen plotted for DPC in \autoref{fig:pc_plot} and for computing connected components in \autoref{fig:perlin_strong}. The raw timing in seconds are presented in \autoref{tab:overview}.
A comparison of speedup and parallel efficiency can be seen in \autoref{fig:pc_speedup} for DPC and in \autoref{fig:perlin_strong_speedup} for computing connected components.

We have seen that the distribution step of path compression does not scale well, as more nodes significantly increases the size of the needed communication. However, the computation of connected components scales well as only few components stretch over multiple ranks and need to be distributed. This shows that the computation of connected components is more dependant on the actual data distribution, while path compression is more independent but can be much slower. 

\begin{figure}[h!]
    \centering
    \includegraphics[width=0.49\columnwidth]{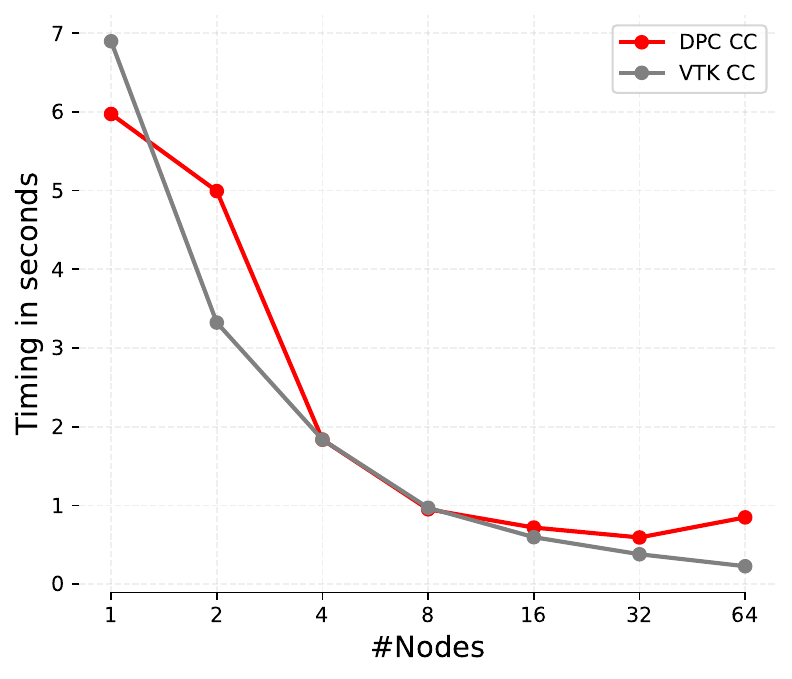}
    \includegraphics[width=0.49\columnwidth]{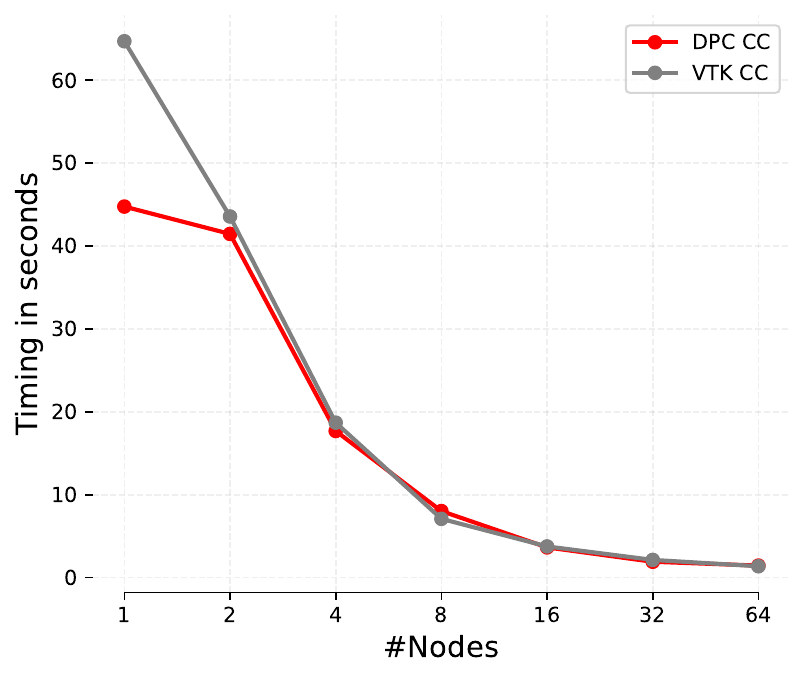}
    \includegraphics[width=0.49\columnwidth]{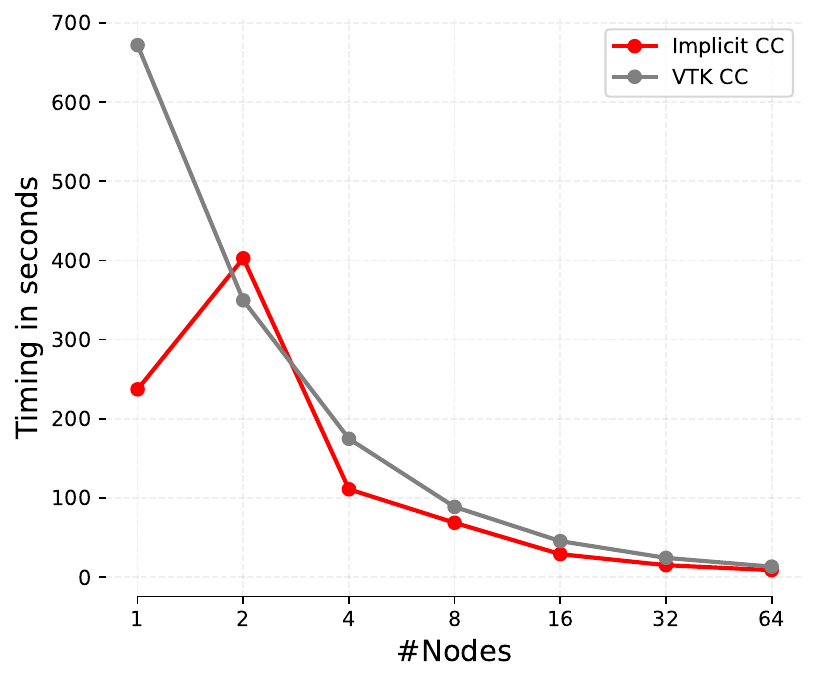}
    \caption{Timing for the strong scaling experiments for computing connected components with DPC (red) and with the VTK Connectivity filter (gray) based on Perlin Noise at a grid size of $512^3$ (upper left), $1024^3$ (upper right) and $2048^3$ (lower) vertices.}
    \label{fig:perlin_strong}
\end{figure}
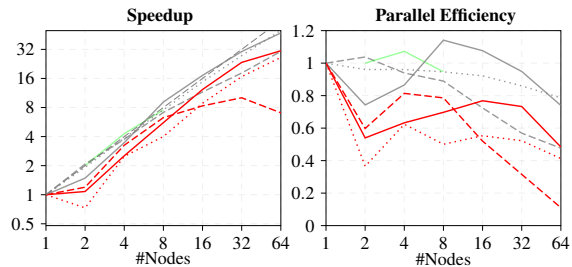
\begin{figure}[h!]
    \centering
    \resizebox{0.9\columnwidth}{!}{\input{figures/speedup_512}}
    \caption{Comparison of parallel speedup and efficiency between computing connected components based on a feature mask implicitly with DPC (red), extracting this feature and explicitly computing connected components with DPC (green, could not be run on all configurations and starting at 2 nodes) and with the VTK Connectivity filter (gray) on Perlin Noise at $512^3$ \rev{(dashed)}, $1024^3$ \rev{(solid)} and $2048^3$ \rev{(dotted)} vertices, thresholded such that the top 10\% of vertices are masked as relevant and are extracted.
    The left plot shows the parallel speedup~(y-axis) defined as the runtime of one node divided by the runtime of $n$ nodes~(x-axis), perfect scalability is marked with the dashed gray line.
    The right plot shows the parallel efficiency~(y-axis) as the speedup divided by the number of nodes~(x-axis).
    The plot shows that the computation of connected components scales well when only few components stretch over multiple ranks and need to be distributed. The smaller grid leads to highly unstable timings at high node counts. 
    }
    \label{fig:perlin_strong_speedup}
\end{figure}

\subsection{Weak Scaling}
\begin{figure}[htb]
    \centering
    \includegraphics[width=\columnwidth]{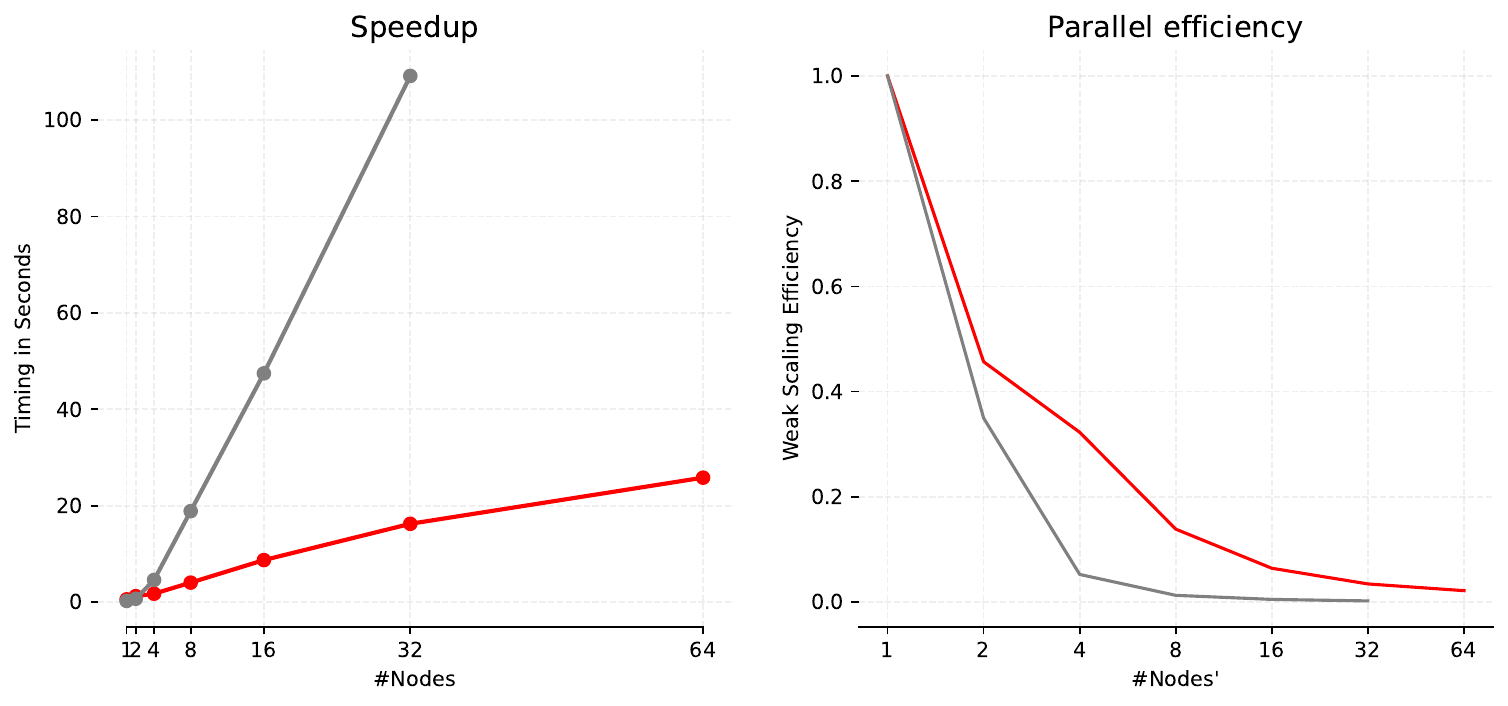}
    \includegraphics[width=\columnwidth]{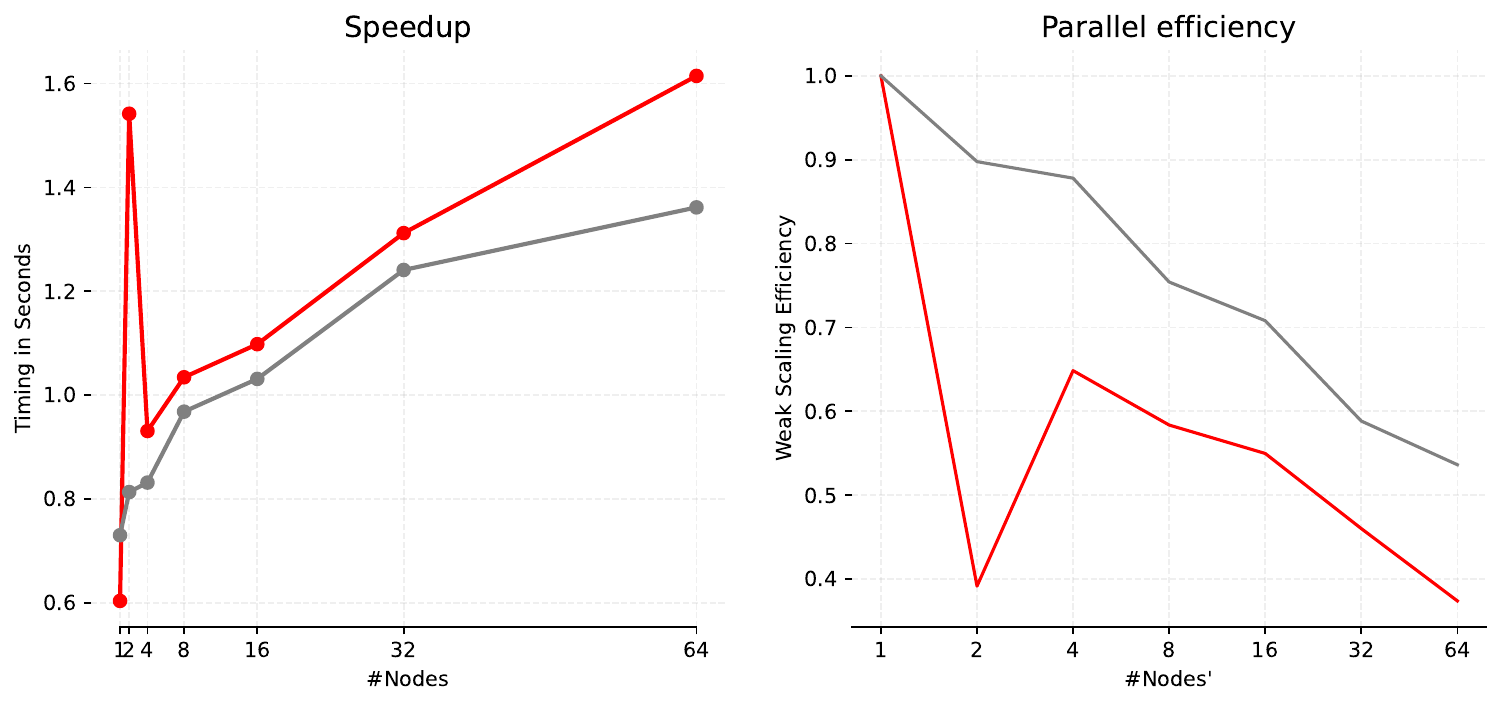}
    \caption{Timing (left) and parallel efficiency (right) for the weak scaling experiments \rev{of DPC (red) and the VTK Connectivity filter (gray)} based on the AT complex (upper) and Perlin Noise (lower) with a low amount of masked vertices and extracted geometry. Perfect weak scaling parallel efficiency would be 1. The VTK based connected components, running on unstructured grids, could not be run on the AT complex for the last configuration, due to memory constraints.}    
    \label{fig:weak}
\end{figure}

We run weak scaling experiments based on Perlin Noise with a high threshold and on the resampled AT complex, the timings can be found in \autoref{tab:weak}. 
\autoref{fig:weak} plots the timings and weak parallel efficiency for the AT complex and Perlin Noise.

\input{tables/weak}

One can see that computing the connected components implicitly is at least as fast as computing it with the VTK connectivity filter and they show very similar scaling behaviour, with the implicit connected components scaling slightly worse for the Perlin Noise and slightly better for the AT complex.
However, Perlin Noise was thresholded with a high scalar value, extracting only little geometry and leaving little actual work to do. 
In contrast, looking at the AT complex, where more geometry was extracted, VTK connectivity failed at the highest node-count / dataset size configuration, due to memory issues.

\subsection{Extraction based on different thresholds}
\input{tables/different_thresholds}

One major advantage of our connected components computation is that it can run \emph{implicitly} on a pre-computed feature mask. 
Therefore we also ran experiments with Perlin Noise of size $1024^3$ and three different computations of features: first we normalize the scalar values and then we either mark everything above 0.9, above 0.5 or above 0.1 as a feature. 
The values of Perlin Noise are normally distributed, so therefore we either mark almost nothing ($\approx 0.06\%$, an unstructured grid with 671 960 vertices and 206 993 cells, for an original grid size of $1024^3$), roughly half of the domain, or almost everything($100\%-\approx 0.06\%$) as a feature.
While this feature masking works easily for our approach, the VTK connectivity filter only works on actively extracted geometry. 
This leads to drastically higher memory usage and possibly worse timing behaviours. 
We present the timings for this in \autoref{tab:different_thresholds}, run on different node counts with a Intel Xeon Gold 6126 and large amounts of memory each (256GB per node for 1,2 and 4, 64GB per node for the rest).
Note, that for the lowest thresholds, most of the configurations could not be run for the VTK connectivity, due to the high memory usage.

While the VTK connectivity is faster than our method for high thresholds (less geometry / less work), we see that it scales worse with increased geometry than our method, with our method being significantly faster at lower thresholds and the connected components being able to be computed even on one node (even when restricting their memory to 64GB). 
In contrast, for the two lower thresholds, the distributed memory of 4 nodes, so 1024GB in total, were needed.

\subsection{Limitations}
It is not only possible to implicitly compute the connected components via DPC, but to extract geometry and then run it on the geometry without a feature mask. 
However, in addition to the previously described problems with the distribution of unstructured grids in TTK, this requires the usage of a triangulation and further increases the memory usage, making it less efficient than the existing VTK connectivity filter.

One of the primary limitations of DPC is its poor scalable efficiency. The global nature of the path compression paradigm necessitates extensive data exchange between nodes. 
As the number of ranks grows, the overhead associated with maintaining global consistency increases significantly, leading to worse performance and reduced scalability. 
This issue becomes particularly pronounced when the size of the dataset grows slower than the number of nodes.

Another limitation is related to the current global communication, which may lead to congestion and redundant work done on the nodes. 
Currently, rank 0 collects the edges for inter-rank connectivity with the specific path compression targets and sends them to all the other ranks, which build up a connectivity graph and path compress it on their own.
Other approaches would be to let rank 0 compress the graph and only send the relevant edges to each rank or to shift to inter-node, neighborhood communication, which would lead to more communication steps with interleaved path compressions, but less work in each communication step.

There may also be ways to further reduce the amount of ghost vertices which need to be sent to minimize communication.
For connected components, it may be feasible to only send a few ghost vertices at the boundaries because they may belong to the same component. 
Including those ghost vertices which only point to other ghost vertices will also reduce the communication sent, but increase the program complexity to account for all edge cases.

\FloatBarrier

%% file: tables/overview.tex
\begin{table*}[htb]
    \centering
\resizebox{\linewidth}{!}{
\begin{tabular}{c !\lv r | r !\lv r !\lv r !\lv r !\lv r !\lv r !\lv r}
\textbf{Size in \#Vertices } & \textbf{Algorithm}                            &                            \textbf{1N} &                            \textbf{2N} &                            \textbf{4N} &                            \textbf{8N}  &                            \textbf{16N}  &                            \textbf{32N}  &                            \textbf{64N}  \\
\midrule

$512^3$ & Segmentation & 11.291 & 20.629 & 1.468 & 2.440 & 3.687 & 4.875 & 6.785 \\

\lh
$1024^3$ & Segmentation & 86.044 & 167.134 & 17.417 & 13.010 & 16.148 & 20.046 & 25.103 \\

\lh
$2048^3$  & Segmentation & 905.872 & 2165.203 & 215.517 & 106.430 & 91.730 & - & - \\

\lh
\midrule
\midrule
\multirow{2}{*}{$512^3$} 
& DPC CC & 5.973 & 4.994 & 1.835 & 0.950 & 0.718 & 0.592 & 0.847 \\
& VTK CC & 6.898 & 3.323 & 1.836 & 0.969 & 0.595 & 0.379 & 0.226 \\

\lh
\multirow{2}{*}{$1024^3$} 
& DPC CC & 44.7504 & 41.451 & 17.683 & 8.023 & 3.640 & 1.908 & 1.446 \\
& VTK CC & 64.692 & 43.553 & 18.686 & 7.086 & 3.756 & 2.131 & 1.367 \\

\lh
\multirow{2}{*}{$2048^3$} & DPC CC & 237.242 & 402.772 & 111.073 & 68.752 & 29.125 & 15.188 & 8.912 \\
& VTK CC & 671.906 & 349.608 & 174.838 & 88.756 & 45.556 & 24.410 & 13.350 \\
\lh
\multirow{2}{*}{$4096^3$} & DPC CC & - & - & - & - & 277.634 & 129.562 & 73.757 \\
& VTK CC & - & - & - & - & 372.957 & 198.001 & 109.066 \\
\end{tabular}
}\vspace{0.5em}
\caption{
    Timing results of our experiments run on Perlin Noise first grouped by dataset size, and then by algorithm, comparing connected component computation using Distributed Path Compression (DPC) and the VTK Connectivity filter.
    The remaining columns show, per node count, the total runtime of the different algorithms in seconds.
    For some datasets the computations ran out of memory or failed because of the previously described reasons and are therefore missing from the table.
}
    \label{tab:overview}
\end{table*}

%% file: figures/speedup_pc.tex

\newcommand{\ssscaleX}{0.6}
\newcommand{\ssscaleY}{0.6}

\tikzstyle{spEdgeR}=[red, line width=0.25mm,opacity=1]
\tikzstyle{spEdgeB}=[blue, line width=0.25mm,opacity=0.8]
\tikzstyle{spEdgeG}=[gray, line width=0.25mm,opacity=0.8]
\tikzstyle{spEdgeGr}=[green, line width=0.25mm,opacity=0.4]

\hspace*{-1.6em}%
\begin{tikzpicture}

\definecolor{brown1672968}{RGB}{167,29,68}
\definecolor{darkcyan26118156}{RGB}{26,118,156}
\definecolor{darkgray176}{RGB}{176,176,176}
\definecolor{gray}{RGB}{128,128,128}
\definecolor{lightgray204}{RGB}{204,204,204}

\begin{groupplot}[group style={group size=2 by 1,horizontal sep=2.5em}]
\nextgroupplot[
log basis x={2},
log basis y={2},
tick align=outside,
tick pos=left,
x grid style={darkgray176,opacity=0.2, dashed},
xlabel={\#Nodes},
xmajorgrids,
xmin=1, xmax=64,
xmode=log,
xtick style={color=black},
y grid style={darkgray176,opacity=0.2, dashed},
ymajorgrids,
ymin=0, ymax=50,
ymode=log,
ytick style={color=black},
log ticks with fixed point,
xscale=\ssscaleX,yscale=\ssscaleY
]
\coordinate (c1) at (rel axis cs:0,1);

\addplot [spEdgeB, dash pattern=on 3.7pt off 1.6pt]
table {
1 1
2 0.5832226526031937
4 8.097009743467463
8 5.383778594110903
16 3.5625176766072406
32 2.6948735565069217
64 1.9360482640635122
};

\addplot [spEdgeB] 
table {%
1.0 1.0
2.0 0.538006814870658
4.0 4.45095876777656
8.0 6.911568024596464
16.0 5.568460490463215
32.0 4.485657986630749
64.0 3.582022069075409
};

\addplot [gray, dash pattern=on 3.7pt off 1.6pt]
table {%
1 1
2 2
4 4
8 8
16 16
32 32
64 64
};

\nextgroupplot[
log basis x={2},
tick align=outside,
tick pos=left,
x grid style={darkgray176,opacity=0.2, dashed},
xlabel={\#Nodes},
xmajorgrids,
xmin=1, xmax=64,
xmode=log,
xtick style={color=black},
log ticks with fixed point,
y grid style={darkgray176,opacity=0.2, dashed},
ymajorgrids,
ymin=0, ymax=2,
ytick style={color=black},
xscale=\ssscaleX,yscale=\ssscaleY
]
\coordinate (c2) at (rel axis cs:1,1);

\addplot [spEdgeB, dash pattern=on 3.7pt off 1.6pt]
table {
1 1
2 0.29161132630159686
4 2.024252435866866
8 0.6729723242638629
16 0.22265735478795254
32 0.0842147986408413
64 0.030250754125992377
};
\addplot [spEdgeB] 
table {%
1.0 1.0
2.0 0.269003407435329
4.0 1.11273969194414
8.0 0.863946003074558
16.0 0.34802878065395093
32.0 0.14017681208221092
64.0 0.05596909482930327
};
\end{groupplot}
    \node[above] at (2,3.4) {\textbf{Speedup}};
    \node[above] at (7,3.4) {\textbf{Parallel Efficiency}};

\end{tikzpicture}

%% file: figures/speedup_512.tex

\newcommand{\ssscaleX}{0.6}
\newcommand{\ssscaleY}{0.6}

\tikzstyle{spEdgeR}=[red, line width=0.25mm,opacity=1]
\tikzstyle{spEdgeB}=[blue, line width=0.25mm,opacity=0.8]
\tikzstyle{spEdgeG}=[gray, line width=0.25mm,opacity=0.8]
\tikzstyle{spEdgeGr}=[green, line width=0.25mm,opacity=0.4]

\hspace*{-1.6em}%
\begin{tikzpicture}

\definecolor{brown1672968}{RGB}{167,29,68}
\definecolor{darkcyan26118156}{RGB}{26,118,156}
\definecolor{darkgray176}{RGB}{176,176,176}
\definecolor{gray}{RGB}{128,128,128}
\definecolor{lightgray204}{RGB}{204,204,204}

\begin{groupplot}[group style={group size=2 by 1,horizontal sep=2.5em}]
\nextgroupplot[
log basis x={2},
log basis y={2},
tick align=outside,
tick pos=left,
x grid style={darkgray176,opacity=0.2, dashed},
xlabel={\#Nodes},
xmajorgrids,
xmin=1, xmax=64,
xmode=log,
xtick style={color=black},
y grid style={darkgray176,opacity=0.2, dashed},
ymajorgrids,
ymin=0, ymax=50,
ymode=log,
ytick style={color=black},
log ticks with fixed point,
xscale=\ssscaleX,yscale=\ssscaleY
]
\coordinate (c1) at (rel axis cs:0,1);
\addplot [spEdgeG, dash pattern=on 3.7pt off 1.6pt]
table {
1 1
2 2.075852308183731
4 3.7576370449163323
8 7.11618892654838
16 11.584179859659562
32 18.20826520459955
64 30.485053551132278
};
\addplot [spEdgeGr]
table {
2 2
4 4.29
8 7.576
};
\addplot [spEdgeR, dash pattern=on 3.7pt off 1.6pt]
table {
1 1
2 1.1961141369643573
4 3.255255585831063
8 6.287783157894737
16 8.31949025069638
32 10.090192567567568
64 7.051580687049936
};
\iftrue
\addplot [spEdgeG]
table {%
1.0 1.0
2.0 1.4853626615847357
4.0 3.4620740091567925
8.0 9.129524148377568
16.0 17.224760389673616
32.0 30.357354859452276
64.0 47.328386641319774
};
\addplot [spEdgeR] 
table {%
1.0 1.0
2.0 1.0795975971629153
4.0 2.5307018039925353
8.0 5.578072312754141
16.0 12.294065934065934
32.0 23.454088050314464
64.0 30.94771784232365
};
\addplot [spEdgeG, dotted] 
table {%
1.0 1.0
2.0 1.9218817151621301
4.0 3.8430153127998588
8.0 7.570263861907543
16.0 14.74899316192058
32.0 27.52582464151619
64.0 50.32910708942333
};
\addplot [spEdgeR, dotted] 
table {%
1.0 1.0
2.0 0.7307972574608326
4.0 2.4949960986531874
8.0 4.006177822373681
16.0 8.849892366218732
32.0 16.714687279252278
64.0 26.3987119899182
};
\fi
\addplot [gray, dash pattern=on 3.7pt off 1.6pt]
table {%
1 1
2 2
4 4
8 8
16 16
32 32
64 64
};

\nextgroupplot[
log basis x={2},
tick align=outside,
tick pos=left,
x grid style={darkgray176,opacity=0.2, dashed},
xlabel={\#Nodes},
xmajorgrids,
xmin=1, xmax=64,
xmode=log,
xtick style={color=black},
log ticks with fixed point,
y grid style={darkgray176,opacity=0.2, dashed},
ymajorgrids,
ymin=0, ymax=1.2,
ytick style={color=black},
xscale=\ssscaleX,yscale=\ssscaleY
]
\coordinate (c2) at (rel axis cs:1,1);
\addplot [spEdgeG, dash pattern=on 3.7pt off 1.6pt]
table {
1 1
2 1.0379261540918654
4 0.9394092612290831
8  0.8895236158185476
16 0.7240112412287226
32 0.5690082876437359
64 0.47632896173644185
};
\addplot [spEdgeGr]
table {
2 1
4 1.0724
8 0.947
};
\addplot [spEdgeR, dash pattern=on 3.7pt off 1.6pt]
table {
1 1
2 0.5980570684821787
4 0.8138138964577657
8 0.7859728947368422
16 0.5199681406685237
32 0.3153185177364865
64 0.11018094823515524
};
\iftrue
\addplot [spEdgeG] 
table {%
1.0 1.0
2.0 0.7426813307923679
4.0 0.8655185022891981
8.0 1.141190518547196
16.0 1.076547524354601
32.0 0.9486673393578836
64.0 0.7395060412706215
};
\addplot [spEdgeR] 
table {%
1.0 1.0
2.0 0.5397987985814576
4.0 0.6326754509981338
8.0 0.6972590390942677
16.0 0.7683791208791209
32.0 0.732940251572327
64.0 0.48355809128630706
};
\addplot [spEdgeG, dotted] 
table {%
1.0 1.0
2.0 0.9609408575810651
4.0 0.9607538281999647
8.0 0.9462829827384429
16.0 0.9218120726200363
32.0 0.860182020047381
64.0 0.7863922982722396
};
\addplot [spEdgeR, dotted] 
table {%
1.0 1.0
2.0 0.3653986287304163
4.0 0.6237490246632968
8.0 0.5007722277967102
16.0 0.5531182728886708
32.0 0.5223339774766337
64.0 0.4124798748424719
};
\fi
\end{groupplot}
    \node[above] at (2,3.4) {\textbf{Speedup}};
    \node[above] at (7,3.4) {\textbf{Parallel Efficiency}};

\end{tikzpicture}

%% file: tables/weak.tex
\begin{table*}[htb]
    \centering
    \resizebox{\linewidth}{!}{%
        \begin{tabular}{c !\lv r | r !\lv r !\lv r !\lv r !\lv r !\lv r !\lv r}
            \toprule
             Dataset & \#Nodes & 1 & 2 & 4 & 8 & 16 & 32 & 64 \\
            \midrule
               \multirow{4}{*}{AT complex} & Segmentation Computation & 1.963 & 8.505 & 1.359 & 2.569 & 5.885 & 11.839 & 26.006 \\
               \cmidrule{2-9}
             & Implicit CC & 0.560 & 1.226 & 1.735 & 4.038 & 8.709 & 16.236 & 25.810 \\
            & Explicit CC & 0.012 & 0.312 & 1.079 & 7.384 & 20.126 & 50.634 & - \\
            & VTK CC & 0.242 & 0.691 & 4.600 & 18.867 & 47.442 & 109.149 & - \\
            \midrule
            \midrule
              \multirow{4}{*}{Perlin Noise} & Segmentation Computation & 1.477 & 5.453 & 0.798 & 2.435 & 5.900 & 10.452 & 25.414 \\
            \cmidrule{2-9}
& Implicit CC & 0.604 & 1.542 & 0.931 & 1.034 & 1.098 & 1.312 & 1.615 \\
& Explicit CC & 0.010 & 0.886 & 0.819 & 1.048 & - & - & - \\
& VTK CC & 0.730 & 0.813 & 0.832 & 0.968 & 1.031 & 1.241 & 1.361 \\
        \end{tabular}
    }
    \caption{Timing for Weak Scaling experiments based on the Adenine Thymine (AT) complex and on Perlin Noise, resampled to different sizes to increase with increased node count, starting from $256^3$ vertices and doubling with every doubled node count, up to $1024^3$ vertices. Algorithms marked with \emph{-} could not be run due to memory constraints.}
    \label{tab:weak}
\end{table*}

%% file: tables/different_thresholds.tex
\begin{table*}[htb!]
    \centering
\resizebox{0.9\linewidth}{!}{
\begin{tabular}{c !\lv r | r !\lv r !\lv r !\lv r !\lv r !\lv r }
\textbf{Size in \#Vertices } & \textbf{Algorithm}                            &                            \textbf{1N} &                            \textbf{2N} &                            \textbf{4N} &                            \textbf{8N}  &                            \textbf{16N}  &                            \textbf{32N}  \\                            
\midrule
\multirow{2}{*}{Top 10\%}
& Implicit CC   & 47.285    & 23.433    & 9.086 & 4.686 & 2.582 & 1.431 \\ 
& VTK CC        & 2.394     & 1.308     & 0.737 & 0.410 & 0.255 & 0.256 \\ 

\lh
\multirow{2}{*}{Top 50\%}
& Implicit CC & 163.410 & 217.288 & 189.417 & 95.898 & 49.360 & 26.345 \\ 
& VTK CC & 675.794 & 396.708 & 210.169 & 94.0211 & 62.176 & 38.642 \\ 

\lh
\multirow{2}{*}{Top 90\%} & Implicit CC & 226.593 & 368.826 & 356.776 & 181.304 & 92.330 & 48.426 \\ 
& VTK CC & - & - & 523.862 & 241.716 & 118.921 & 78.366 \\ 
\lh
\end{tabular}
}\vspace{0.5em}
\caption{
    Timing results of implicit connected components and VTK connectivity run on Perlin Noise of size $1024^3$ with different feature thresholds, extracting the Top 10\%, Top 50\% and Top 90\% of the values.
    We see that, while being faster for small data, for larger data / more extraction, VTK connectivity drastically slows down for larger extraction. Additionally, most of the more intensive workflows could not be run at all due to memory constraints.
}
    \label{tab:different_thresholds}
\end{table*}

%% file: sections/conclusion.tex
\section{Conclusion}
\rev{We described an adaption of a well-scaling parallel algorithm for computing Morse-Smale segmentations to a distributed setting, additionally using it as a base to efficiently compute connected components on distributed structured and unstructured grids.
Furthermore, we provide an implementation in TTK, which is open source and integrated in the widely-used ParaView visualization environment, and conduct a series of scaling experiments on large-scale datasets in distributed environments.}

Future work will focus on further optimizing communication patterns and exploring additional applications of our methods in various scientific domains. 
The integration of our algorithms into TTK provides a solid foundation for continued development and application of topological data analysis tools in distributed computing environments.
Some experiments could not be run due to shortcomings of VTK and TTK.
We aim to further investigate these problems, address them and run more comprehensive benchmarks, with different scientific visualization datasets and feature masks, and comparing different communication approaches.
Currently, resolving path compression on the ghost vertices requires global communication.
It is possible to change this to only need rank neighborhood communication, at the cost of additional compression and communication steps. 
Additionally, there is a clear trade-off between dataset size and number of ranks. 
In contrast to shared memory parallelism where more threads will often give at least \emph{some} performance improvement, with distributed computations, more ranks can significantly worsen performance, due to the communication increase.
Therefore it is paramount to use as many threads as possible per rank and only as many ranks as are actually needed for the dataset.
We plan to conduct more thorough experiments on these trade-offs to find out when this is worth it.
Additionally, there are ways to additionally minimize the amount of ghost vertices which need to be communicated, such as compressing paths along ghost vertices, but only if they point to other ghost vertices.
These optimizations possibly introduce many edge cases and therefore need to be cautiously evaluated.

Having the ability to compute the ascending and descending segmentations in a distributed setting allows us to efficiently compute the \emph{extremum graph} used in the merge tree algorithm \mbox{ExTreeM}~\cite{lukasczyk_extreem_2024}. 
As there is currently no distributed merge tree algorithm available in TTK, extending \mbox{ExTreeM} to a distributed setting would allow for much more sophisticated analysis on large datasets.